\documentclass{aa}  

\usepackage{graphicx}
\usepackage{txfonts}
\begin{document}

   \title{RR~Lyrae stars with variable mean magnitudes}

   %\subtitle{Subtitle}

%%%%%%%%%%%%%%%%%%%%%%%%%%%%%%%%%%%%%%%%
% Please do not include ORCIDs next to author names.
% Only ORCIDs authenticated by individual authors in EDP Sciences editorial system will be taken into account.
% ORCIDs included here will be removed.
%%%%%%%%%%%%%%%%%%%%%%%%%%%%%%%%%%%%%%%%

   \author{G. Hajdu\inst{1}
      \and J. Jurcsik\inst{2}\fnmsep%\thanks{Shows the usage of elements in the author field}
      \and M. Catelan\inst{3,4}\fnmsep%\thanks{Shows the usage of elements in the author field}
      \and G. Pietrzy\'{n}ski\inst{1,5}\fnmsep%\thanks{Shows the usage of elements in the author field}
      \and V. Hocd\'{e}\inst{1}\fnmsep%\thanks{Shows the usage of elements in the author field}
      \and I. Soszy\'{n}ski\inst{6}\fnmsep%\thanks{Shows the usage of elements in the author field}
      \and A. Udalski\inst{6}\fnmsep%\thanks{Shows the usage of elements in the author field}
      \and C.-U. Lee\inst{7}\fnmsep%\thanks{Shows the usage of elements in the author field}
      \and D.-J. Kim\inst{7}\fnmsep%\thanks{Shows the usage of elements in the author field}
        }

   \institute{Nicolaus Copernicus Astronomica Center, Bartycka 18, 00-629 Warsaw, Poland\\
             \email{ghajdu@camk.edu.pl}
             %\thanks{Shows the usage of elements in the author field}
            \and Konkoly Observatory, HUN-REN Research Centre for Astronomy and Earth Sciences, Konkoly-Thege Mikl\'{o}s \'{u}t 15-17,
            Budapest H-1121, Hungary 
            \and Instituto de Astrof\'isica, Pontificia Universidad Cat\'olica de Chile, Av. Vicu\~na Mackenna 4860, Macul 7820436, Santiago, Chile
            \and Millenium Institute of Astrophysics, Nuncio Monse\~nor Sotero Sanz 100, Providencia, Santiago, Chile
            \and Departmento de Astronom\'{i}a, Universidad de Concepci\'{o}n, Casilla 160-C, Concepci\'{o}n, Chile 
            \and Astronomical Observatory, University of Warsaw, Al.~Ujazdowskie~4, Warsaw 00-478, Poland
            \and Korea Astronomy and Space Science Institute, Daejeon 34055, Republic of Korea
            }

   \date{Received September 30, 20XX}

\abstract
{A number of RR~Lyrae stars show variable mean magnitudes in the OGLE survey light curves of the Galactic bulge.
Hitherto this phenomenon was not studied, as it was generally assumed to be related to problems
with the photometry.}
{We investigate whether the mean magnitude variability of RR~Lyrae variables is due to genuine astrophysical phenomena.}
{We make use of the extended, and in many cases overlapping, light curves from multiple microlensing surveys, to study RR~Lyrae stars with
apparent mean-magnitude variations.
A modified Fourier-series based fitting method is introduced to analyze the light curves showing mean-magnitude variations.
Data from infrared surveys are also used to construct spectral energy distributions (SEDs).
}
{72 stars are presented where the mean-magnitude variations are most probably of genuine astrophysical origin, and not the result
of problems with the photometry. The ratio of variation between the $V$ and $I$ bands is compatible with variable extinction
by dust in most cases, but no infrared excess is detected in the SEDs. The occurrence rate of the phenomenon, after correcting for selection
effects, is $\sim0.9\%$ among RR~Lyrae variables in the OGLE bulge fields.}
{}


   \maketitle
   \nolinenumbers
\section{Introduction}

RR~Lyrae (RRL) variable stars are widely used tracers and probes of the properties
of old stellar populations, both in the Milky Way
\citep{2017ApJ...850...96H,2019MNRAS.487.3270P,2020AJ....159..270K,2022MNRAS.513.1958W,2024MNRAS.527.8973C}
and in other galaxies of the Local Group
\citep{1995AJ....110.2166M,2003ApJ...588L..85C,2012ApJ...756..121M,2013ApJ...765...71C,2017AcA....67....1J,2017AJ....154..263K,2018MNRAS.478.4590T,2022ApJ...938..101S}.

The Optical Gravitational Lensing Experiment (OGLE; \citealt{2015AcA....65....1U}), as part of its regular survey observations, monitors the
Galactic bulge at a high cadence, incidentally providing light curves for tens of thousands of RRL stars in this region \citep{2014AcA....64..177S}.
We have recently undertaken a search for RRL in binary systems using this treasure trove of data \citep{2021ApJ...915...50H},
using the light-travel time effect (LTTE; \citealt{1952ApJ...116..211I}).
During our search, we have also visually inspected the light curves of each of the analyzed $\sim27,\!000$ fundamental-mode
RRL stars (RRab subclass), and noticed that some of them show long-term variability
in their mean brightness. In previous studies, these were either ignored, or assumed to be due to photometric trends,
and subsequently removed \citep{2017MNRAS.465.4074P,2018MNRAS.480.1229N}.

Due to the importance of RRL variables as stellar population tracers and standard candles for distance determination, it is crucial to investigate
possible sources of bias affecting their properties. Therefore, we have decided to perform a systematic revision of mean-magnitude changes
in the same sample of stars using photometry from OGLE, as well as other surveys. The results are presented in this article, which is organized
as follows: Section~\ref{sec:data} presents the utilized data, Section~\ref{sec:analysis} describes the light curve analysis procedure,
in Section~\ref{sec:sample} the properties of the sample are discussed, and finally in Section~\ref{sec:discussion}
we put the results in context and hypothesize on the possible origin of the mean-magnitude changes.

\begin{figure*}[t!]
    \sidecaption
    \includegraphics[width=12cm]{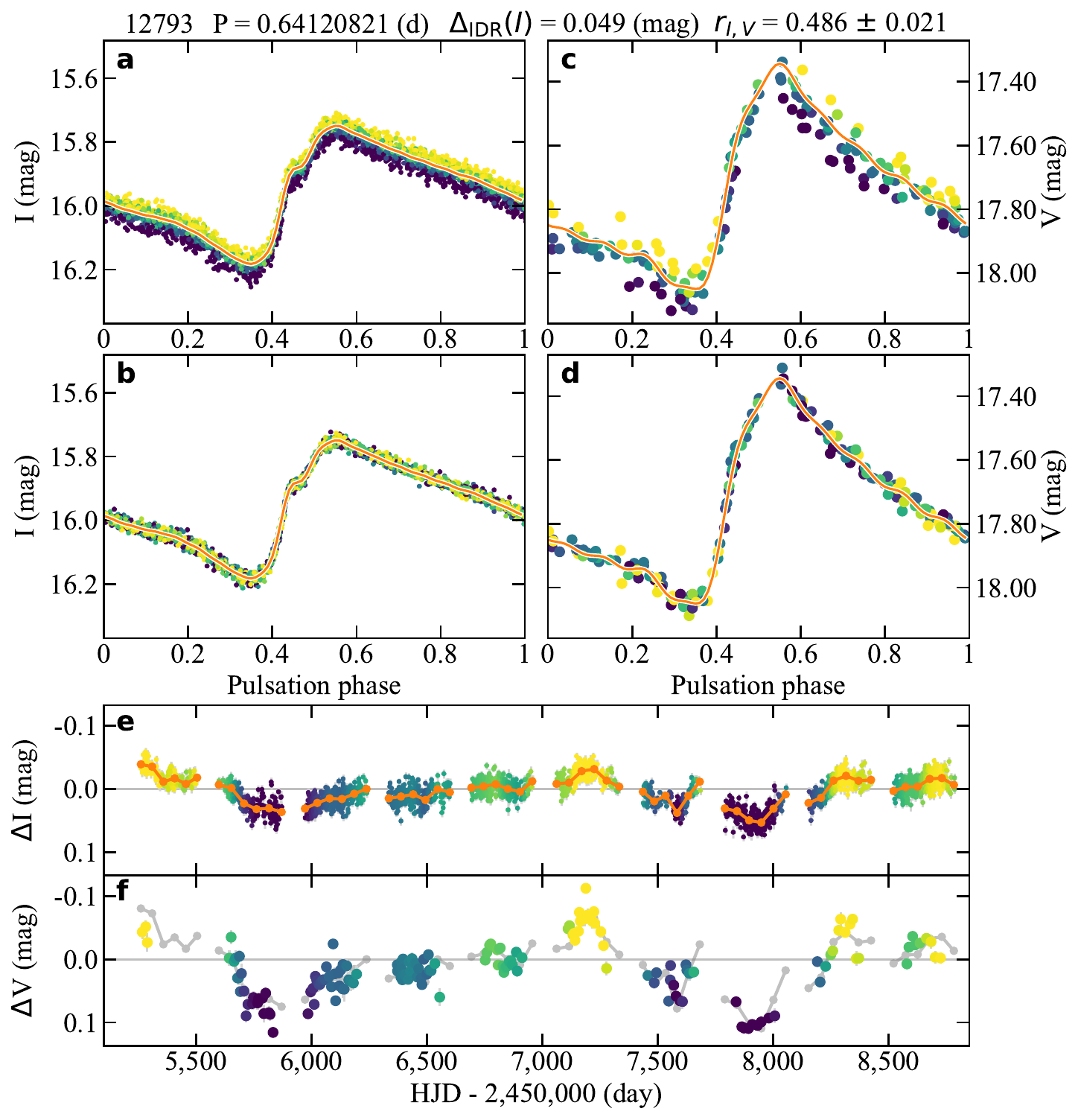}
    \caption{Summary plot for one of the analyzed stars.
    {\it Panel a:} the original OGLE $I$-band light curve folded with the period of pulsation, with the corresponding Fourier fit overlaid.
    {\it Panel b:} same as ({\it a}), but with the long-term changes removed.
    {\it Panel c:} same as ({\it a}), but for the $V$-band data.
    {\it Panel d:} same as ({\it b}), but for the $V$-band data.
    {\it Panel e:} the $I$-band data after the subtraction of the Fourier light-curve model.
    The orange dots and lines show the fit of the long-term brightness changes.
    {\it Panel f:} same as ({\it e}), but for the $V$-band light curve.
    All light-curve points are colored according to the estimated mean-magnitude values at the times of the observations,
    with lighter and darker points corresponding to brighter and fainter epochs, respectively.
    Above the panels, the OGLE ID, the pulsation period, the amplitude ($\Delta_\mathrm{IDR}(I)$) and
    the amplitude ratio ($r_{I,V}=A(I)/A(V)$) of the mean-magnitude changes are given.
    {\it The complete set of summary plots is available as online material.}
    }\label{fig1}
\end{figure*}

\section{Utilized data} \label{sec:data}

The main analysis was performed using the Johnson-Kron-Cousins $V$ and $I$-band time series photometric measurements
of OGLE's Galactic bulge RRL sample~\citep{2014AcA....64..177S}, obtained during the survey's fourth phase (OGLE-IV; \citealt{2015AcA....65....1U}).
The currently public data were extended for the analyzed objects with unpublished data from the 2018 and 2019 observing seasons.
Furthermore, we also extracted from the OGLE internal catalogs the light curves of stars within 10 arcseconds of RRL stars on
our short list of candidates, and used those to assess possible photometric contaminations (see Section~\ref{sec:analysis}).
Light curves available from the Massive Compact Halo Object (MACHO; \citealt{1993Natur.365..621A}), Exp\'{e}rience pour la Recherche d'Objets Sombres phase~2
(EROS-2; \citealt{2008A&A...481..673T}), Korea Microlensing Telescope Network (KMTNet; \citealt{2016JKAS...49...37K}),
Microlensing Observations in Astrophysics (MOA; \citealt{2008ExA....22...51S}) surveys, and the third phase of the OGLE project (OGLE-III; \citealt{2011AcA....61....1S})
were also revised during the analysis, in order to search for independent evidence of mean-magnitude changes similar to those detected in the OGLE-IV photometry.
Additional details about the utilized optical photometry, including the transformation of the KMTNet and MOA differential fluxes onto a magnitude scale, are given in 
Appendix~\ref{sec:data2}.

We also utilize proper motions taken from the {\em Gaia} Data Release~3  catalog (DR3; \citealt{2023A&A...674A...1G}).
However, in the dense environment of the bulge, the catalog is incomplete and proper motions have relatively large errors for the
highly reddened RRL variables. 
Furthermore, formal errors in the {\em Gaia} DR3 catalog might be underestimated by a factor of up to 4 in the most dense bulge fields
\citep{2023A&A...677A.185L}.
Therefore, for our final sample, we have replaced the {\em Gaia} DR3 proper motions for
RRab stars on our list with the values from
the pre-release version of the OGLE-Uranus proper motion catalog (Udalski et al., in prep.). 
The OGLE-Uranus catalog has on average smaller formal errors, especially for the most reddened ($V-I>2$) variables.

During the construction of spectral energy distributions (SEDs; Section~\ref{sec:sed}), we also make use of the near-infrared (near-IR) $JHK_\mathrm{S}$ light curves
obtained by the Vista Variables in the V\'{i}a L\'{a}ctea (VVV; \citealt{2010NewA...15..433M}) survey, as well as 
the mid-infrared flux measurements provided by the Galactic Legacy Mid-Plane Survey Extraordinaire (GLIMPSE; \citealt{2009PASP..121..213C}) family of surveys.
\begin{table*}[!ht]
\centering
\caption{\label{tab:properties} Analysis summary of the final sample of RR Lyrae with mean-magnitude changes}
\begin{tabular}{c@{\extracolsep{2.0em}}c@{\extracolsep{2.0em}}c@{\extracolsep{2.0em}}c@{\extracolsep{2.0em}}r@{\extracolsep{0em}}lr|ccrrc}
\hline \hline
& \multicolumn{6}{@{}c@{}|}{Preset parameters}& \multicolumn{4}{@{}c@{}}{Derived parameters} \\
\hline
ID & Period    & $O-C$ & $N_{\mathrm{bp,s}}$ & \multicolumn{2}{c}{$N_{I}$}  & \multicolumn{1}{c|}{$N_{V}$} & $\Delta_\mathrm{IDR}(I)$ & $r_{I,V}$ &  $\mathcal{F}_{I,V}$ & $\Delta_\mathrm{AIC}$ & LC\\
   & (days) & & & & & & (mag) \\
\hline
00663 & 0.61682289 & C     & 5                 &  1,153 & (1)    &  67(2)      & 0.034 & $0.505\pm0.089$ &    5,3              &  22.2         & 1\\
00835 & 0.61922115 & C     & 5                 &  1,210 & (0)    &  69(2)      & 0.042 & $0.782\pm0.136$ &   10,3              &  23.1         & 1\\
00853 & 0.55825120 & P     & 5                 &  1,210 & (0)    &  69(1)      & 0.054 & $0.583\pm0.089$ &   13,5              &  30.0         & 1\\
\vdots & \vdots    & \vdots& \vdots            & \vdots &         & \multicolumn{1}{c|}{\vdots} & \vdots &   \vdots & \multicolumn{1}{c}{\vdots} & \multicolumn{1}{c}{\vdots} & \vdots \\
32226 & 0.56209850 & B     & 5                 & 16,690 & (14)   & 218(1)      & 0.029 & $0.391\pm0.033$ &   30,7              &  97.3         & 2\\
33665 & 0.73443630 & P     & 5                 &  8,147 & (5)    & 213(2)      & 0.031 & $0.354\pm0.026$ &   29,5              & 134.1         & 3\\
34373 & 0.32525898 & C     & 8                 &  2,236 & (1)    & 186(0)      & 0.259 & $0.386\pm0.010$ &    5,1              & 430.2         & 4\\
\hline
\end{tabular}
\tablefoot{
The first column provides the OGLE identifier in the format OGLE-BLG-RRLYR-$ID$.
The second to sixth columns give the parameters used during the fitting process:
the second column  gives the pulsation period adopted for the analysis;
the third the type of phase correction applied to the data, based on the $O-C$ curves
(C: constant period, no correction applied; P: parabola subtracted;
B: binary solution subtracted);
the fourth the adopted number of breakpoints per observing season for the analysis;
the fifth and sixth the number of the data points and the number of manually removed points in parentheses
for the $I$ and $V$ bands, respectively.
The seventh to tenth columns give the derived parameters.
The seventh column gives the interdecile ranges calculated on the $I$-band mean magnitude nodes;
the eight column lists the derived amplitude ratio of the mean magnitude changes, $A(I)/A(V)$, with uncertain values marked with an asterisk;
the ninth the adopted Fourier orders of our final light-curve solutions in the $I$ and $V$ bands, respectively;
the tenth the AIC difference between our varying mean-magnitude model and a simple Fourier series fit for the $V$ band.
The eleventh column encodes the qualitative appearance of the long-term changes in the light curves as:
(1) mostly monotonic decrease or increase;
(2) pronounced intervals of both decreasing and increasing behavior;
(3) quasi-periodic variation on a timescale shorter than the length of the data;
(4) complex variations.}
\end{table*}

\section{Search for RRL variables with changing mean-magnitudes} \label{sec:analysis}

During our initial search for mean-magnitude changes in RRL stars, we have visually inspected the OGLE-IV light curves from
the 2010-2017 observing seasons for the same sample of RRL stars as in our search for binary RRL \citep{2021ApJ...915...50H}.
A total of $\sim250$ stars were selected for this long list of candidates, only omitting stars where it was
obvious from the light curves that the mean-magnitude changes were caused by photometric problems (e.g., the mean magnitude
showed a clear correlation with the pulsation amplitude).

The RRL light curves in the long-term OGLE photometry can suffer from various biases. They can blend together with other
variable stars, might be affected by light scattered within the camera from very bright stars in the field,
or trends can arise from problems with the image subtraction photometry caused by their changing position on the sky over time
(for stars with high proper motions).
Therefore, the OGLE-IV $I$- and $V$-band light curves were inspected for any suspicious trends associated
with these effects. Time series data from other surveys were also revised during this analysis,
and a few stars with completely stable light curves in these were also removed.
A particularly notable group of stars in the field of the globular cluster NGC~6441 was removed altogether, as their light curves all showed
the same general trend and a common breakpoint in their mean-magnitude behavior in the OGLE-IV light curves, whereas the
OGLE-III light curves were completely stable. In this case, we suspect that the foreground star G~Sco ($V\sim3.2$\,mag) is the source of the
contamination in the OGLE-IV data.

All in all, more than half of the stars from the initial long list were removed in this step, resulting in a short list
of $\sim100$ variables. These were further revised and their light curves were fit multiple times using both traditional Fourier series,
as well as a modified Fourier series allowing for changing mean magnitudes (see Appendix~\ref{sec:fourier} for a description of the
latter method).
Figure~\ref{fig1} shows a typical example for the analysis with mean-magnitude changes.
The solutions were iterated and analyzed to identify and remove the last few dubious cases. The light curves of all sources detected by
OGLE-IV within 10 arcseconds of these RRL were also extracted and analyzed.
The light curves of a handful of RRL variables were clearly contaminated by other nearby variable stars, and were therefore removed from the list.
Furthermore, we require the OGLE-IV $I$ and $V$ band light curves to show the same general trends in the mean magnitudes of the variables.
To this end, we calculate the difference in the Akaike Information Criterion (AIC; \citealt{1974ITAC...19..716A}) between
the traditional Fourier series and the mean-magnitude-changing Fourier series solutions in the $V$ band ($\Delta_\mathrm{AIC}$).
A few stars with $\Delta_\mathrm{AIC}$ of less than 20 were removed at this step\footnote{A $\Delta_\mathrm{AIC}$ value
of 20 corresponds to a relative likelihood between the two models of
$\exp (\Delta_\mathrm{AIC}/2) \sim 22,\!000$. For the commonly adopted $\Delta_\mathrm{AIC}$ value of 10, the relative
likelihood between two models is only $\sim 148$.}.
Unfortunately, some good candidates have too few $V$-band detections, or were not at all detected by OGLE-IV in the $V$ band
(due to the large amount of extinction), and these stars were also removed.
Other candidates display annual trends in their photometry, which are clearly visible in the changes
of the mean magnitudes. Variables where the only clear reason for the mean-magnitude changes are these
annual trends were removed from the sample.
In contrast, we chose to retain variables where the long-term mean-magnitude changes are completely
explained by our light-curve modeling for both bands, and other sources of photometry also confirm this effect,
despite their light curves also having annual trends
(such as OGLE-BLG-RRLYR-01591, -08752, and -11931).

\begin{figure*}[t!]
    \centering
    \includegraphics[width=0.98\hsize]{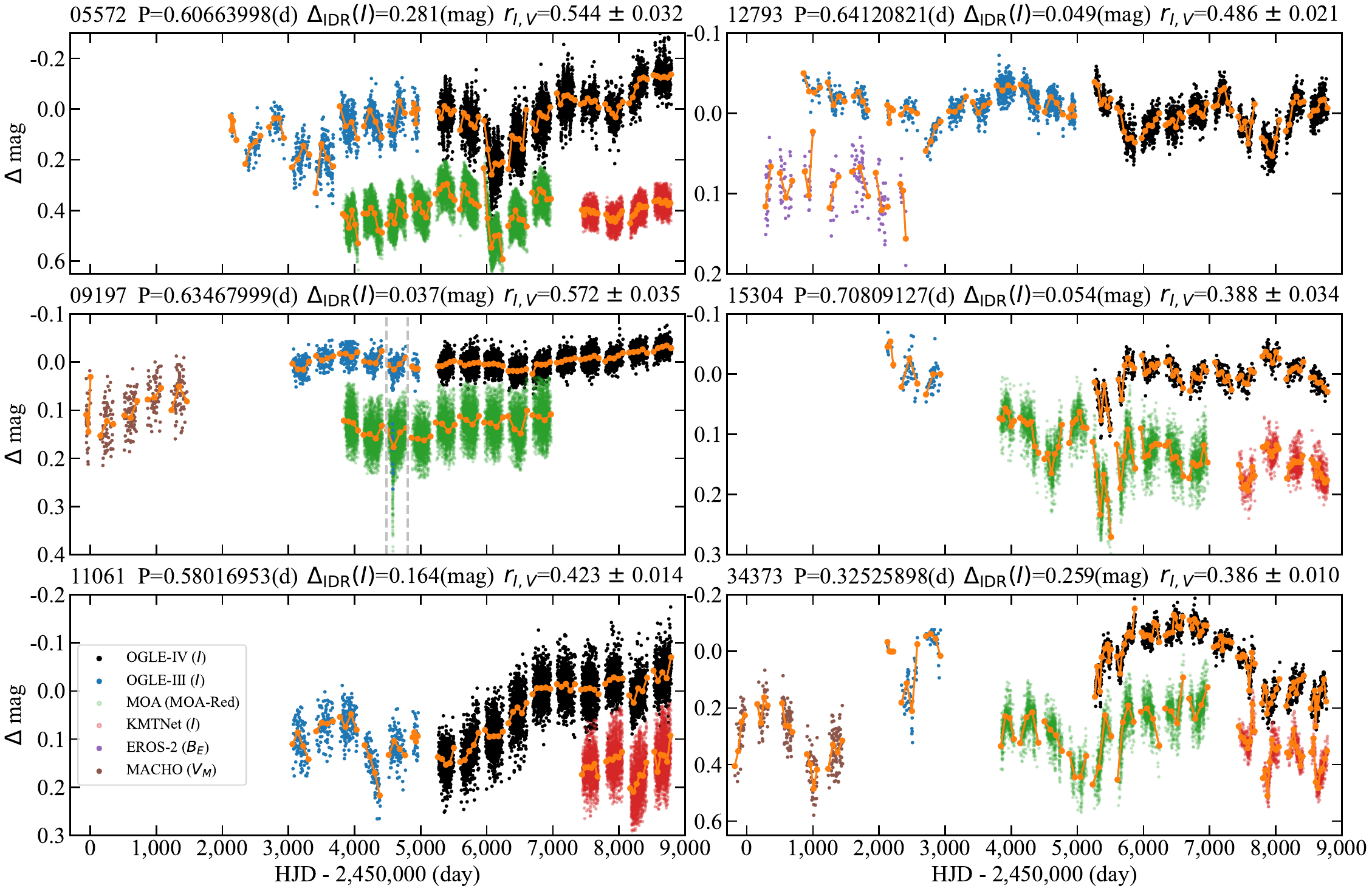}
    \caption{
    Comparison of RRL mean-magnitude changes between different surveys.
    Each panel shows the OGLE-IV, OGLE-III, MACHO, EROS-2, MOA, and KMTNet light curves (black, blue, brown, purple, green, and red dots, respectively;
    see their descriptions in Appendix~\ref{sec:data2}) after the subtraction of (separate) Fourier light-curve models for each data set.
    Light curves are offset by arbitrary amounts for clarity.
    The overlaid orange dots and lines show the mean-magnitude change modeling according to the light-curve fits, similarly to panel ({\it e}) of Fig.~\ref{fig1}.
    Above the panels, the OGLE ID, the pulsation period, the amplitude of the mean-magnitude changes,
    and the measured amplitude ratio $r_{I,V}=A(I)/A(V)$ are given for the corresponding RRL, based on the OGLE-IV data.
    }\label{fig:composite}
\end{figure*}

\section{Properties of the final sample} \label{sec:sample}

After the final revision of the short list of candidates, we were left with 71 RRab stars, which all display variable mean magnitudes
in both their $I$- and $V$-band photometry. While we have not yet performed a dedicated search among first-overtone RRL stars (RRc subclass) for this effect,
one such variable, OGLE-BLG-RRLYR-34373, has a remark on its variable mean magnitude in the OGLE bulge RRL catalog \citep{2014AcA....64..177S}.
We concluded, after analyzing its light curve, that it shows the same phenomenon as the RRab variables, therefore it was added to the sample considered here,
for a total of 72 RRL stars on this final list.
The properties of these variables are summarized in Table~\ref{tab:properties}.
We note that during the fitting process, a few outlying light curve points have been removed manually for most variables,
and the corresponding number of points is also given in the table. The amplitude of the mean-magnitude changes is quantified by the
interdecile range in the $I$-band ($\Delta_\mathrm{IDR}(I)$) calculated on the mean magnitudes ($m_{0,z}$) at the times of each adopted breakpoint ($t_z$)
of the modified Fourier fit, as described in Appendix~\ref{sec:fourier}. Furthermore, the amplitude ratio of the mean-magnitude changes, defined as $r_{I,V}=A(I)/A(V)$, is also listed, alongside 
the Fourier orders selected by our method in the $I$ and $V$ bands, and the difference in AIC values calculated for the $V$-band
solutions. Finally, the appearance of the mean-magnitude changes in the OGLE-IV light curves is classified into four categories:
(1) mostly monotonic increase or decrease;
(2) both increasing and decreasing intervals at different times;
(3) quasi-periodic appearance;
(4) exhibiting complex variations.

The long-term behavior of the mean-magnitude changes differs between stars, as shown in Figure~\ref{fig:composite}.
The OGLE-IV data (black dots) show that the majority of variables display the smoothest
mean-magnitude behavior near maximum brightness. In contrast, stars with a large amplitude in the mean-magnitude
variations typically display sharp, asymmetric drops, mostly in their dimmer light-curve phases, although some
variables with lower amplitudes can behave similarly as well (see the {\it bottom-right} and {\it top-right} panels of Figure~\ref{fig:composite} for respective examples).
Light curves from surveys overlapping with OGLE observations show an excellent agreement in the shapes
of the mean-magnitude behavior for the majority of cases. As these data are independent from each other, this eliminates
the possibility that the observed mean-magnitude changes are caused by problems with the OGLE-IV photometry.
Earlier surveys usually have lower photometric accuracy and fewer data points, but they also show the same kind of light-curve
behavior as more recent data. In a few cases, quasi-periodic patterns also appear on long time scales. OGLE-BLG-RRLYR-12793
({\it top-right} panel of Figure~\ref{fig:composite}) shows a possibly periodic (or quasi-periodic) behavior with a tentative period of $\sim2500$\,d or $\sim5000$\,d;
OGLE-BLG-RRLYR-34373 ({\it bottom-right} panel) similarly shows a possible periodicity on a timescale of $\sim4000$\,d or $\sim8000$\,d.
Information on the presence of detectable mean-magnitude changes in the MACHO, EROS-2, OGLE-III, MOA, and KMTNet light curves is given in Appendix~\ref{sec:otherinfo}
for all stars on our final list.

We note that while our fitting works well for most stars, some (i.e., OGLE-BLG-RRLYR-02401, -04444, and -10084)
show minor deviations in the mean-magnitude changes in the $V$ band, when compared to the $I$ band. It is possible that these differences are caused
by unaccounted-for biases in the photometry, or alternatively, could be real differences between the behavior of the phenomenon between the $V$ and $I$ bands.

\subsection{Light-curve properties}\label{sec:lcparams}

\begin{figure}[t!]
    \centering
    \includegraphics[width=\hsize]{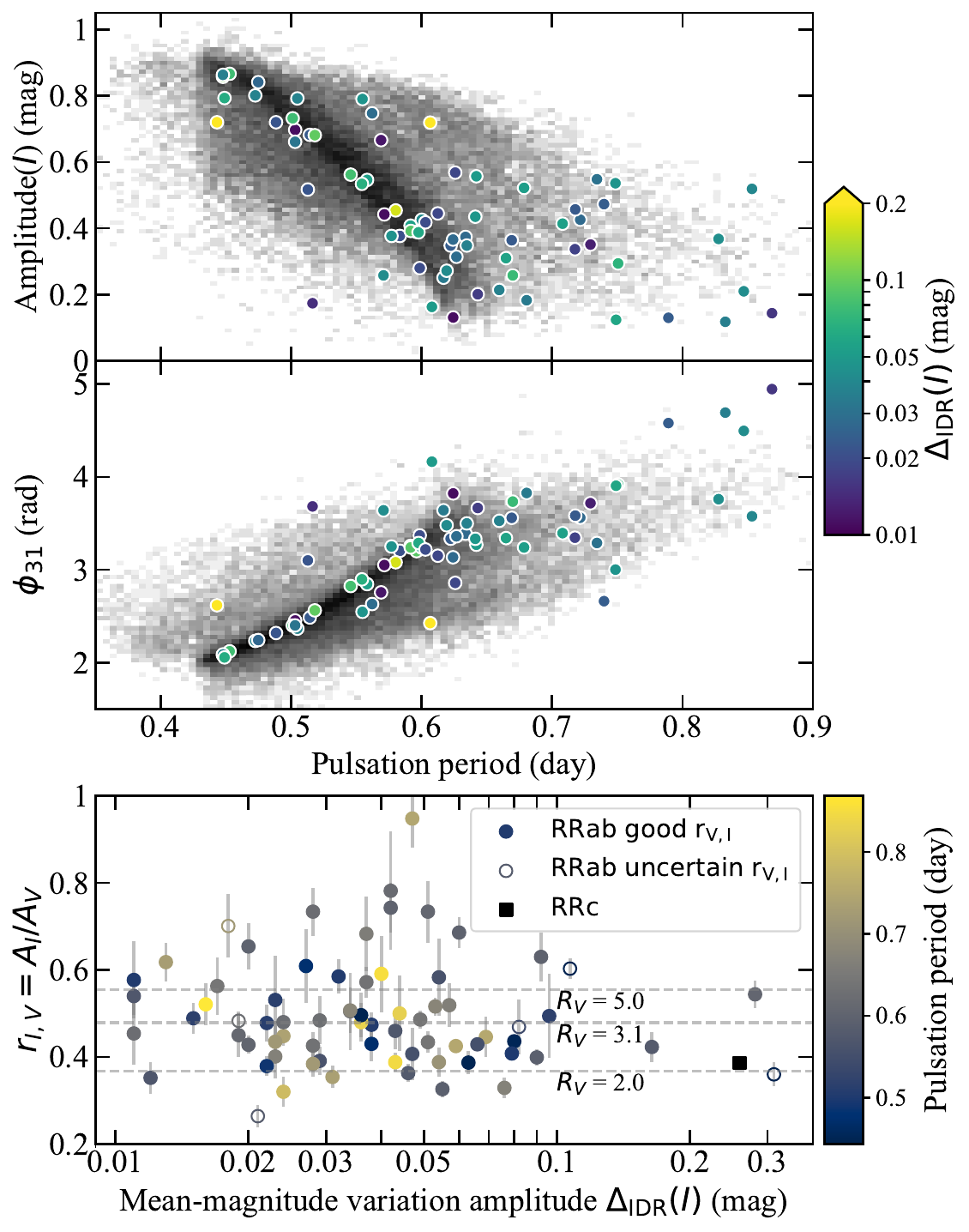}
    \caption{
    {\it Top panel:} period-amplitude (Bailey) diagram of the OGLE bulge RRab sample (grey 2D histogram, shown on a logarithmic scale) and the
    mean-magnitude-changing RRL variables circles colored according to the measured $I$-band mean-magnitude changes.
    {\it Middle panel:} the distribution of the epoch-independent phase differences ($\phi_{31}$) of the RRab stars.
    {\it Bottom panel:} the dependency of the amplitude ratio $r_{I,V}$ on the amplitude of the $I$-band mean-magnitude changes.
    RRab stars are marked with circles, and are colored according to their pulsation periods, while the only RRc star is marked with a black square.
    The horizontal dashed lines show the expected extinction ratios for different $R_{V}$ values of the standard interstellar extinction law \citep{1989ApJ...345..245C}.
    }\label{fig:statistics}
\end{figure}

The light-curve shapes of RRL stars give information about their physical parameters \citep{1996A&A...312..111J,2020MNRAS.491.4752B},
which in turn inform us about their host stellar populations. The {\it top} and {\it middle} panels of Figure~\ref{fig:statistics} compare the distribution of
pulsation amplitudes and the light-curve shape parameter $\phi_{31}$ \citep{1982ApJ...261..586S} of our sample to that of the OGLE survey toward
the Galactic bulge. About half of our sample lay on top of the main ridge of the dominant bulge RRL population \citep{2020AcA....70..121P} on
both panels (the exact number is between 30 and 40, depending on its assumed width). The remainder have light-curve parameters mostly typical of
RRL stars in Oosterhoff~II type systems \citep{catelan2015}, suggesting that they have lower metallicities on average. Some might
be halo interlopers, which make up between 8 and 25 percent \citep{2019MNRAS.487.3270P,2020AJ....159..270K} of the RRL variables observed in the Galactic bulge.

The {\it bottom} panel of Figure~\ref{fig:statistics} compares the amplitude ratio of the mean-magnitude changes in the
two bands ($I,V$) and its total amplitude in
the $I$ band ($\Delta_\mathrm{IDR}(I)$). If the changes are caused by intervening dust, then the amplitude ratio can be
interpreted as an extinction ratio between the two bands. These are compatible with dust composed of submm-sized
particles, as shown by the comparison to the standard interstellar extinction curve \citep{1989ApJ...345..245C}.
However, we do note that there are a few RRL variables with
amplitude ratios more in line with a nearly flat ``grey'' extinction, which can be
caused by a substantially larger average grain size \citep{2001ApJ...548..296W}.
Alternatively, if the changes are interpreted as intrinsic variations of RRL stars,
the amplitude ratios are also comparable to variability caused by stellar spots in red giants.
Specifically, using the OGLE sample of rotating variables toward the Galactic bulge \citep{2024AcA....74....1I},
we calculate 10th percentile, median, and 90th percentile values of the amplitude ratio $A_I/A_V$ of
0.31, 0.52, and 0.79, respectively, for stars with $P_\mathrm{rot}>20$\,d.

\begin{figure}[t!]
    \centering
    \includegraphics[width=\hsize]{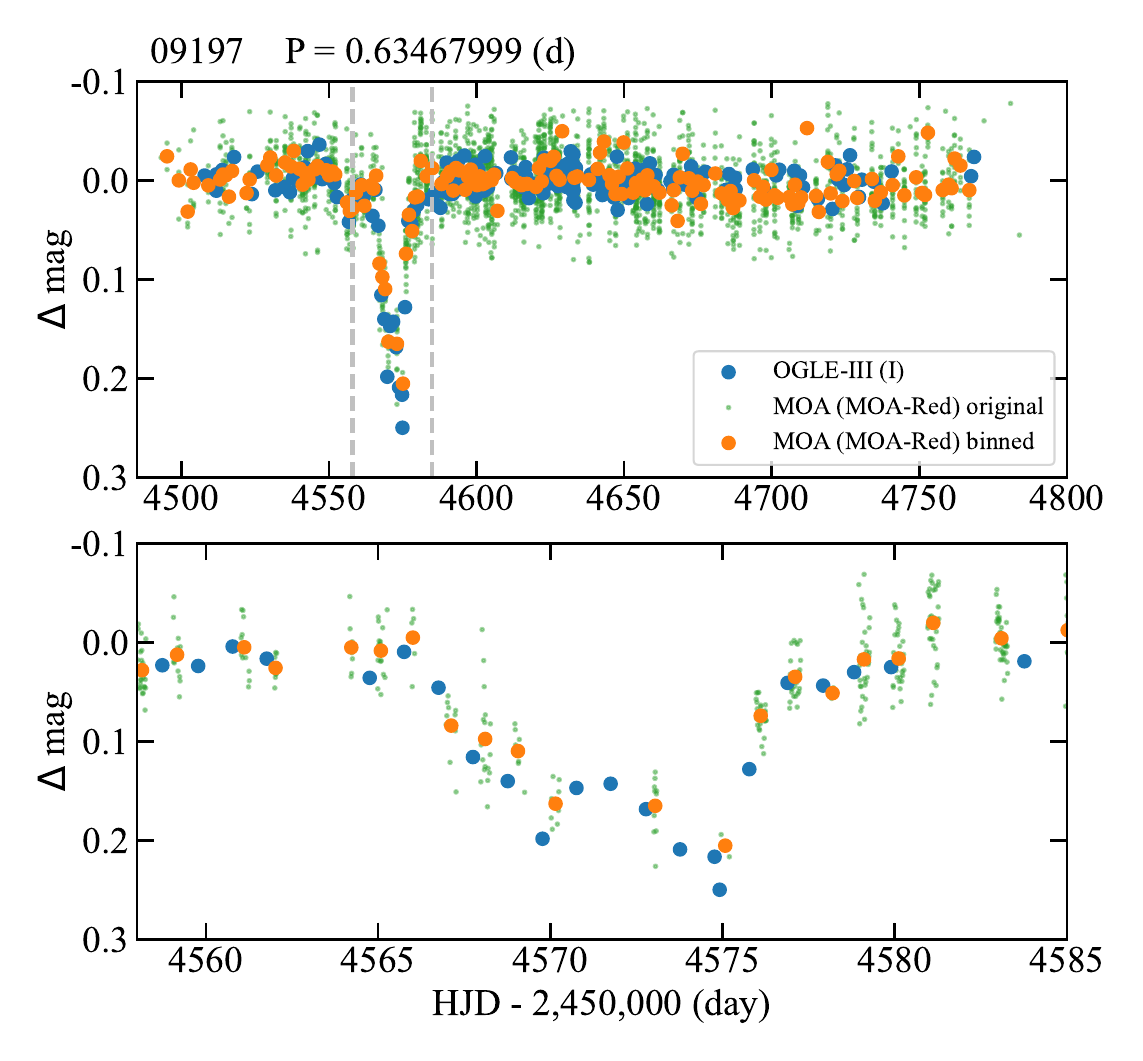}
    \caption{
    The eclipse-like event of OGLE-BLG-RRLYR-09197.
    {\it Top:} The residual light curve of the 2008 observational season (both pulsation and long-term mean-magnitude changes are subtracted).
    OGLE and MOA observations are shown by blue and green dots, respectively. For clarity, daily average magnitudes for the MOA data are also shown
    with orange points.
    {\it Bottom:} Same as the {\it top} panel, but showing the event between the limits marked by dashed vertical grey lines.
    }\label{fig:09197}
\end{figure}

One of the RRL stars showing mean-magnitude changes, OGLE-BLG-RRLYR-09197, has experienced a single eclipse-like fading
during the third phase of the OGLE project \citep{2011AcA....61....1S}. The observing season of the event is marked with vertical dashed lines on
the {\it left-middle} panel of Figure~\ref{fig:composite}. This event is by far the sharpest drop in brightness detected in any of the stars presented here.
Figure~\ref{fig:09197} shows the data from 2008, after the removal of both the pulsation and the long-term mean-magnitude change signals. While few
light curve points from OGLE cover this event, the contemporaneous MOA data unambiguously confirms
it\footnote{\citet{2011AcA....61....1S} reported two more eclipse-like events in the OGLE-III light curves of RRL variables: OGLE-BLG-RRLYR-03593 and -11361.
Unfortunately, to our best knowledge they were not observed during these events by any of the other bulge microlensing surveys,
therefore we cannot provide independent confirmation for them.}.
This quite asymmetric event lasted $\sim12$\,days, with a slow ingress, quick egress, and a local maximum near the center of the event.
This shape is reminiscent of eclipses produced by objects surrounded by rings and/or disks, such as:
the Be star surrounded by at least two circumstellar disks, eclipsing the peculiar W~Virginis variable OGLE-LMC-T2CEP-211 \citep{2018ApJ...868...30P};
the disk-shrouded low-mass companion of a B~star in the eclipsing binary OGLE-LMC-ECL-11893 \citep{2014ApJ...797....6S};
the dark disk surrounding the invisible companion of a Be star in the binary EE~Cep \citep{2012A&A...544A..53G};
and the occulting dark disk in the binary ELHC~10 \citep{2016MNRAS.457.1675G}.

The eclipse event shown in Figure~\ref{fig:09197} makes OGLE-BLG-RRLYR-09197 the best candidate for an RRL variable in an eclipsing binary system;
however, no similar eclipse has been observed before or after this singular event, suggesting a very long orbital period.

\subsection{The incidence rate}\label{sec:incidence}

An estimation of the true incidence rate of the mean-magnitude changes of RRL stars is complicated by the fact
that the sampling of the OGLE light curves strongly depends on the position of the star in the sky. In general,
OGLE-IV fields with high stellar densities in the Galactic bulge regions are observed more frequently, due to the
higher occurrence rate of microlensing events (see Fig.~15 in \citealt{2015AcA....65....1U}).

\begin{figure}[t!]
    \centering
    \includegraphics[width=\hsize]{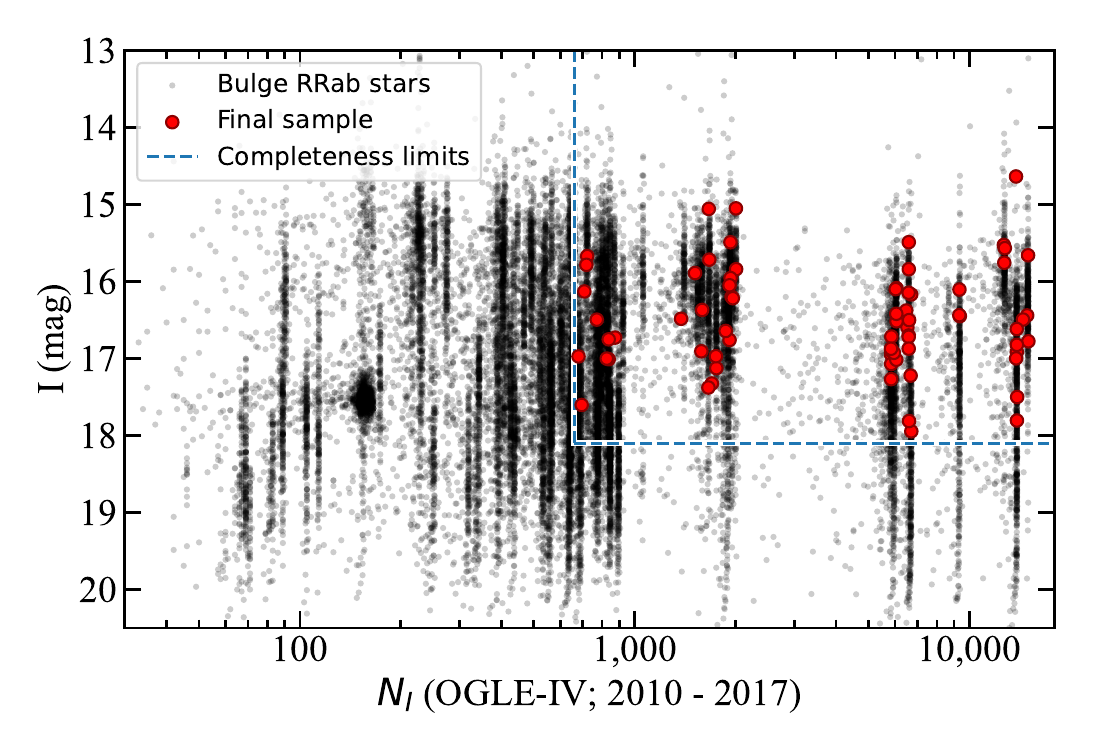}
    \caption{The distribution of the available $I$-band data points and mean magnitudes in the OGLE survey for
    RRab stars (black points).
    Stars of our final sample are marked with red circles. The dashed blue lines denote the region used to calculate the
    incidence rate of the mean-magnitude changing effect in RRab stars.}
    \label{fig:completeness}
\end{figure}

Figure~\ref{fig:completeness} shows the distribution of $I$-band magnitudes as a function of the
number of available $I$-band data points (during the OGLE-IV 2010-2017 seasons) of the
analyzed sample of stars.
Stars in our sample are all relatively bright with $I\lesssim17.95$\,mag and have at least 680 data points.
There are a total of 11,449 stars in the initial sample if we adopt $I<$\,18.1\,mag
and $N_{I} > 660$ as the absolute detection limits. Then, an initial estimate for the
incidence rate of the mean-magnitude changes is 71/11,449\,$\sim$\,0.6$\%$.

An additional complication for detecting the mean-magnitude changes is the presence of the Blazhko effect.
The Blazhko effect modulates the light curve shapes and amplitudes in a large fraction of RR~Lyrae stars with periods of a few days to
multiple years, and with amplitudes ranging from millimagnitudes to multiple tenths of a magnitude \citep{2016pas..conf...22S}.
In our experience, the detection of mean-magnitude changes, which have typical amplitudes below 0.1\,mag in the $I$ band (see the {\it bottom} panel
of Figure~\ref{fig:statistics}), is significantly complicated by its presence.
Assuming that these two effects operate independently from one another,
this can be roughly corrected for by taking into account the fraction of Blazhko stars in the sample
and the occurrence rate of the Blazhko effect, similarly to our investigation of RRL binarity with the
LTTE \citep{2021ApJ...915...50H}.
In the final sample, there are only 10 Blazhko stars ($14\%$; see Appendix~\ref{sec:otherinfo}). The incidence rate of the Blazhko effect in the
OGLE Galactic bulge RRab sample is at least $40\%$ \citep{2017MNRAS.466.2602P}. Given the 61 RRab stars in the
final sample with no Blazhko effect detected, we would expect to detect $\sim40$ stars showing the Blazhko effect
and mean-magnitude variations at the same time (instead of 10), for a total of $\sim101$ detections.
Taking into account the data quantity and brightness cuts, this increases the incidence rate to $\sim\,0.9\%$, which
can be adopted as a lower limit for the incidence rate for RRab stars presenting real changes in their mean magnitudes.
Given this relatively low occurrence rate, we surmise that the mean-magnitude changes reported here have a negligible
effect on the utility of RR~Lyrae stars as standard candles.

\subsection{Sky distribution, color-magnitude diagram and proper motions} \label{sec:properties}

\begin{figure*}[t!]
    \sidecaption
    \includegraphics[width=12cm]{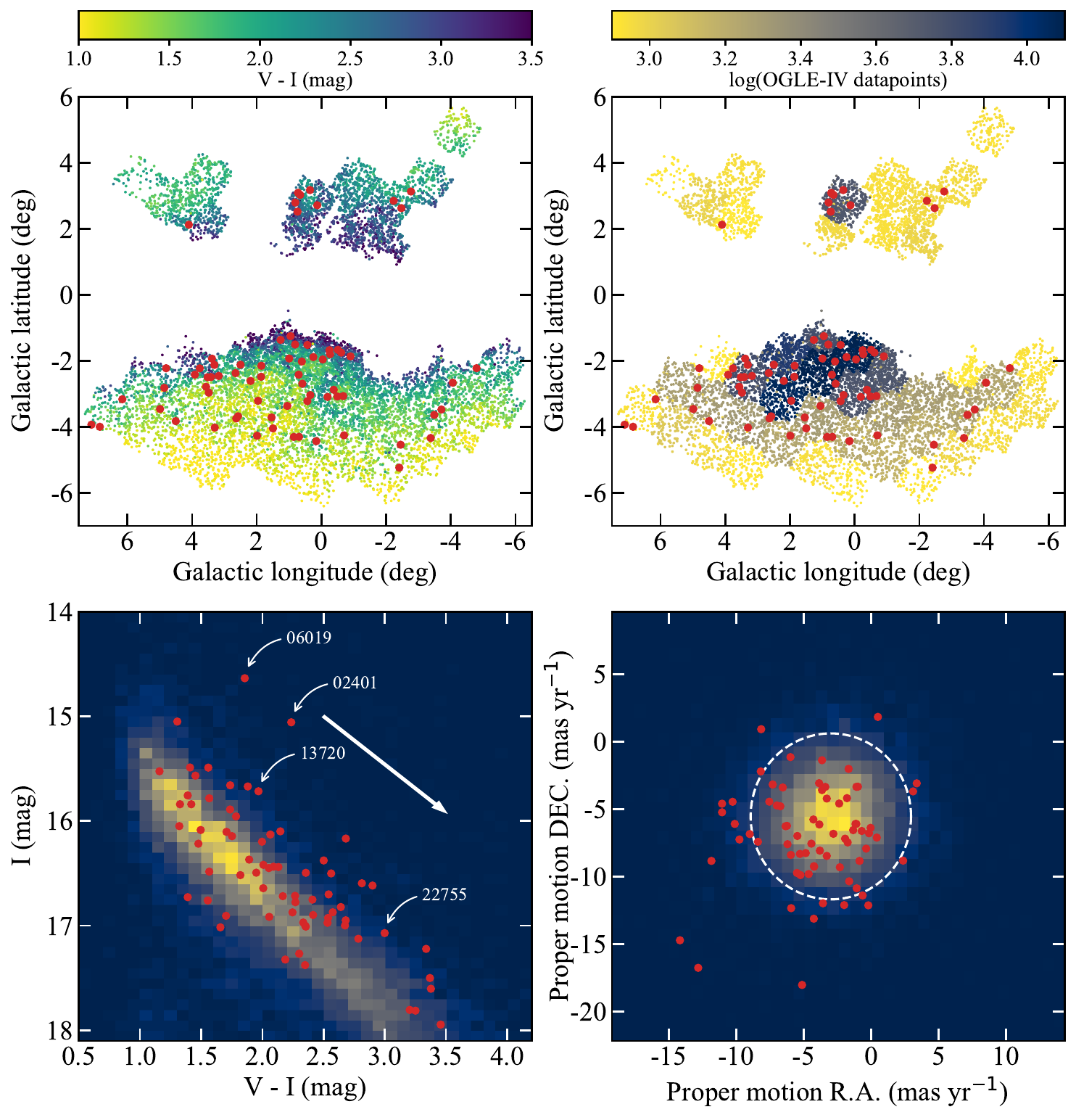}
    \caption{{Comparison of properties of the 71~RRab stars in our sample (red dots) with the stars used for the 
    incidence rate calculation (see Figure~\ref{fig:completeness}).}
    {\it Top left:} the distribution of RRab stars in Galactic coordinates. Colors correspond to the $V-I$ color of each RRab star.
    {\it Top right:} same as {\it top left}, but with the colors corresponding to the number of available $I$-band data points.
    {\it Bottom left:} the $I$ vs. $V-I$ color-magnitude diagram of the RRab stars. The white arrow denotes the reddening vector for 1 magnitude of color excess for
    $R_V=3.1$ \citep{1989ApJ...345..245C}.
    {\it Bottom right:} the proper motion distribution of the sample. The dashed white ellipse denotes the two standard deviation distance from
    the sample median.}
    \label{fig:statistics2}
\end{figure*}

The distribution of stars in our final sample might give us additional hints about the cause of the mean-magnitude changes.
The {\it top panels} of Figure~\ref{fig:statistics2} show the sky distribution of the sample reported here (red dots),
compared to the distribution of RRL stars in the OGLE-IV sample which pass the detectability limits ($I<$\,18.1\,mag and $N_{I} > 660$)
adopted in Section~\ref{sec:incidence}. On the {\it top-left} panel, the RRL variables are colored according to their $V-I$ colors (as a proxy for the reddening),
while on the {\it top-right} panel they are colored according to the number of available OGLE-IV $I$-band data points.
In the southern bulge fields the number density of both our sample and that of the RRab variables in general
steadily increases toward the Galactic plane, and this increase stops only when the extinction pushes the RRL magnitudes beyond our adopted cutoff of $18.1$\,mag.
The {\it top-right} panel also clearly shows that stars from our sample are preferentially located in regions with the largest amount of photometric points.

The {\it bottom-left} panel of Figure~\ref{fig:statistics2} shows the color-magnitude diagram of the final sample,
while the OGLE-IV sample (after applying the detectability cuts) is represented by the
linearly scaled 2D histogram. Curiously, the distribution of our sample does not follow that of the general bulge sample, with our stars
being preferentially closer, and in some cases farther, than the Galactic bulge.
Two stars in our sample (OGLE-BLG-RRLYR-02401 and -06019, marked with arrows) are apparently closer than 4\,kpc, i.e., less than half the distance of the Galactic center.
The difference between the two samples is further corroborated by the distributions of proper motions shown on the {\it bottom-right} panel.
The peak of the distribution of the complete sample (shown again by the 2D histogram scaled linearly) corresponds to the mean motion of bulge stars.
The white dashed ellipse shows the $2\sigma$
distance from the sample median (where the standard deviations were measured using the robust median absolute deviations). As can be seen, a
sizable fraction of stars are quite far from the median of the total sample. To put this into numbers, only $30\%$ and $69\%$ of the stars of the
mean-magnitude-changing RRL stars reported here are within 1$\sigma$ and 2$\sigma$ distance from the sample median, respectively
(compared to the expected $39\%$ and $86\%$, respectively). 
Therefore, the {\it bottom} panels of Figure~\ref{fig:statistics} hint at a difference between the general
(bulge-dominated; \citealt{2019MNRAS.487.3270P,2020AJ....159..270K}) RRL sample and the RRL variables showing mean-magnitude changes,
similarly to the {\it top} and {\it middle} panels of Figure~\ref{fig:statistics}, showing the light-curve parameters of the RRab stars.

In order to correctly assign every RRL of our final sample to its parent stellar population (i.e., Galactic bulge, halo, or disk),
a complete dynamical analysis is necessary. Unfortunately, most of these stars currently lack a radial velocity measurement, preventing
us from performing said analysis. Furthermore, due to the mean-magnitude changes themselves, a judicious application of period-luminosity-metallicity
relationships will be necessary to correctly estimate their distances.

\subsection{Transiting mass estimate and spectral energy distributions} \label{sec:sed}

Under the assumption
that the mean-magnitude changes are caused by variable extinction due to circumstellar dust,
a minimum mass of the material transiting the disk of the RRL stars can be estimated.
We do so by adopting an orbital distance and geometry, as well as a specific
dust mass-to-extinction ratio (see details in Appendix~\ref{sec:dustmass}). This leads to lower limits of
$\sim 1.3 \times 10^{18}$~kg, $\sim 6.7 \times 10^{17}$~kg, and $\sim 1.1 \times 10^{18}$~kg, 
for the RRL variables with the highest inferred extinction amounts, namely,
OGLE-BLG-RRLYR-05572, -11061, and -34373, shown on the {\it top-left, bottom-left} and {\it bottom-right} panels of Fig.~\ref{fig:composite}, respectively.
For comparison, the mass of a 90\,km diameter asteroid with a density of $\sim3$\,g\,cm$^{-3}$
(which is typical of S-type asteroids; \citealt{2021A&A...654A..56V}) is $\sim1.1\times10^{18}$\,kg, i.e., the same order of magnitude
as the inferred dust masses. To put this number into context, this is $\sim5.4$\,million times smaller than the
mass of the Earth.

Circumstellar material can be detected around stars by the infrared excess
in their spectral energy distributions (SEDs), with its strength and shape depending on the mass, temperature, geometry,
and chemical properties of the dust present.
As these features are generally the strongest in the mid- and far-infrared wavelengths, we have searched all infrared archives
for the longest-wavelength observations available to include in our SEDs.
Unfortunately, the large distance, high backgrounds, and crowding
resulted in only three stars being detected in the $8 \, \mu {\rm m}$ band in the GLIMPSE surveys by the {\em Spitzer} space telescope \citep{2004ApJS..154....1W}.
None of our stars were detected by surveys at longer wavelengths.

We construct our SEDs for the three stars which were detected in the $8 \, \mu {\rm m}$ band in the GLIMPSE surveys.
The optical and near-IR parts of the observed SEDs were constructed using the pulsation-averaged magnitudes of the stars.
In the Johnson-Kron-Cousins $V$ and $I$ bands, the values provided by the OGLE-IV catalog were adopted.
In the near-IR $JHK_\mathrm{S}$ bands, the VVV survey light curves were fit using a principal component-based fitting
method \citep{2018ApJ...857...55H}, after correcting the original VVV light curves for known issues affecting their
photometric zero points \citep{2020ExA....49..217H}.
The $VIJHK_\mathrm{S}$ magnitudes were converted to fluxes using photometric zero points provided by the
SVO Filter Profile Service \citep{2020sea..confE.182R}, and $5\%$ measurement uncertainties were assumed for all bands.
For the mid-IR part of the SED, the GLIMPSE flux measurements in the four mid-IR bands were adopted alongside their quoted errors.

\begin{figure}[t!]
    \centering
    \includegraphics[width=0.9\hsize]{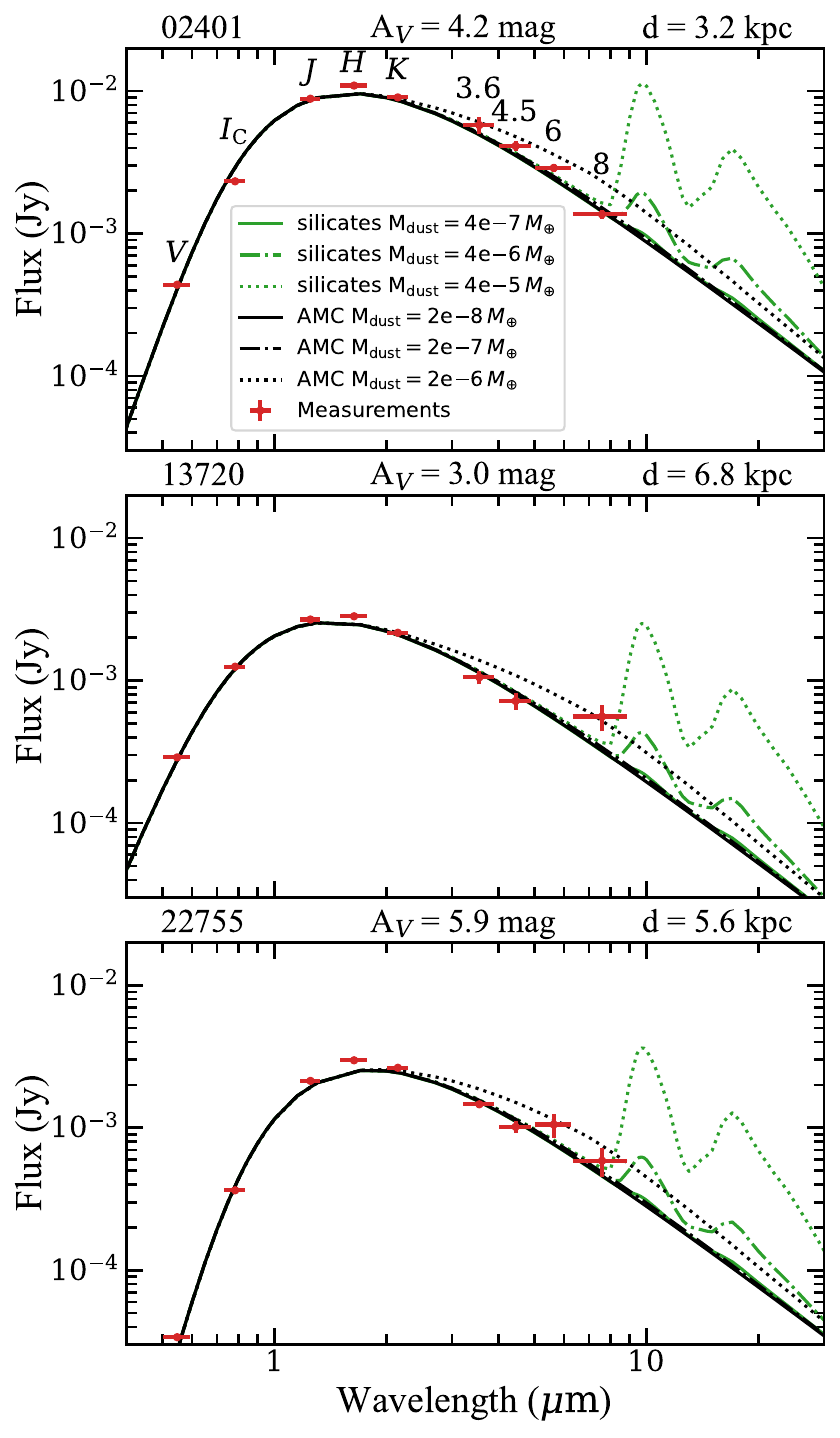}
    \caption{Spectral energy distributions of RRL with $8\mu m$ photometry available.
    On each panel the measured fluxes of a star are marked with red dots. The vertical and horizontal bars mark
    the measurement uncertainties and the effective widths of the photometric bands, respectively.
    The black and green curves correspond to {\tt DUSTY} models with circumstellar envelopes made of amorphous carbon and
    silicate dust, respectively. The continuous, dash-dotted, and dotted lines show progressively higher amounts of dust.
    Above each panel, the OGLE ID, and the adopted values of the $V$-band extinction and distance are given.
    The locations of these three stars are marked by arrows on the color-magnitude diagram of the final sample
    ({\it bottom-left} panel of Fig.~\ref{fig:composite}).
    }
    \label{fig:seds}
\end{figure}

The expected infrared excess was modeled using the radiative transfer code {\tt DUSTY} \citep{1999ascl.soft11001I}.
For the stellar parameters we adopted typical values for RRL variables of $L = 50\,L_\odot$,
$R = 5.3\,R_\odot$, and $T_\mathrm{eff} = 6200\,{\rm K}$ (similar to those of the prototype of the class; \citealt{2010A&A...519A..64K}),
and the stellar SED was approximated as a black body.
The models were constructed for two different dust chemical compositions, one typical for olivine with a 40-60\% magnesium-to-iron ratio,
and one for amorphous carbon \citep{1988ioch.rept...22H}.
The grains were assumed to be spherical, and to follow a standard size distribution \citep{1977ApJ...217..425M}.
The gas-to-dust mass ratio was kept at 140 for all models.
The models were calculated assuming spherical symmetry, with the internal shell temperature set to 800\,K, resulting in an
internal shell radius of $\sim4$\,AU for all models. The thickness of the shell was set to be $10\%$ of the internal shell radius.
For both compositions, a set of models was calculated by increasing the amount of $I$-band extinction from the shell itself in increments of $10\times$.
The total dust mass also scales with this factor between the models.

In order to compare the model SEDs to the flux measurements of the RRL variables, the former were reddened according to the
\citet{2007ApJ...663..320F} interstellar reddening law, using the {\tt extinction} Python package \citep{2016zndo....804967B},
while the distance to the star with the simulated SED was changed simultaneously, until a fair match was reached for
the optical and near-IR fluxes.
As can be seen in Fig.~\ref{fig:seds}, 
our theoretical SEDs match the data well.
It is also obvious from the $3.6$ and $4.5 \, \mu {\rm m}$  fluxes that the observations are incompatible with
dust made of large amounts of exclusively amorphous carbon particles, and that the currently available observations are insensitive to
even large quantities of silicate dust (with the green dotted line showing the model corresponding to dust with $\sim200\times$ the mass
than inferred for the transiting material).
We assess that future observations in wavelengths longer than $8\mu m$ might reveal the silicate emission feature.
However, we also note that the detection of the $10 \, \mu {\rm m}$  silicate emission could be obscured by the absorption of the same feature
by interstellar matter \citep{1984MNRAS.208..481R}, which is non-trivial to correct for, even for comparatively close stars
with low extinction \citep{2020A&A...633A..47H}.

\subsection{Connection to binarity} \label{sec:binaryconnection}

The deep eclipse event discussed in Section~\ref{sec:lcparams} raises the possibility that the observed mean-magnitude changes
are caused by circumstellar matter, as is the case for binary post-asymptotic giant branch (post-AGB) stars \citep{1991A&A...242..433W,1995A&A...293L..25V}.
Establishing a correlation between the presence of binarity and the mean-magnitude changes might lend support to this hypothesis.
Of the 71 RRab stars in our final sample, five are binary candidates detected through the LTTE signal.
Two of these systems (OGLE-BLG-RRLYR-08215 and -08752) have estimated minimum companion masses \citep{2021ApJ...915...50H}
close to the expected mass of old white dwarfs ($\sim 0.6 \, {\rm M}_\odot$; \citealt{2018ApJ...866...21C}), meaning that they could be in such a system with a
co-planar circumbinary disk close to an inclination of 90\,degrees. However, the other candidate companions have much smaller minimum masses,
which can be reconciled if their disks are in a polar configuration with respect to the orbital plane of the binary, as in the case of
the post-AGB star AC~Her \citep{2023ApJ...957L..28M}.

In order to assess the possible connection between the binarity of RRL stars and the detected mean-magnitude changes,
we can estimate the probability of finding stars showing both behaviors in their light curves simultaneously.
In our previous work, we detected 87 binary candidates among RRab stars using OGLE photometry \citep{2021ApJ...915...50H}.
During our analysis, one more variable was found to possess a strong LTTE signal (see Appendix~\ref{sec:newbin}), for a total of 88 such systems.
Out of these, 86 are within the detectability limits shown in Figure~\ref{fig:completeness}. 
Assuming that the binarity and the mean-magnitude changes are unrelated, the probability of randomly selecting (at least) five
common stars from the two samples (the 86 binary candidates, and the 71 mean-magnitude changing RRab variables) out of
11,499 possible objects is $\sim0.02\%$; selecting at least four is $\sim0.2\%$ and at least three is $\sim1.6\%$.
The one-tailed Z test suggests that the $\sim0.02\%$ probability is indicative that the two samples are correlated at a level of $\sim3.5\sigma$,
which we deem significant enough to mention, but do not consider definitive proof yet.
We note that detecting RRL stars in binary systems with the LTTE is a challenging task, due to the long binary periods, limited data, and
the pulsation of RRL stars themselves. In the case of the mean-magnitude-changing RRL stars, photometric uncertainties
are increased by the detected variable extinction, as well. Therefore, a more careful analysis with more data could possibly uncover more
binary candidates in the sample of stars reported here, increasing the significance of the suggested connection between the two phenomena.

\section{Discussion} \label{sec:discussion}

To our knowledge, similar variations in the mean magnitudes have never been reported or predicted for RRL stars in the literature before.
In this section, we discuss the possible causes of the observed mean-magnitude changes in light of the properties of the final sample
(Section~\ref{sec:sample}).

\subsection{Circumstellar/circumbinary material around RR~Lyrae} \label{sec:insystem}

Arguably the simplest explanation for the observed light curves (Figs.~\ref{fig1} and \ref{fig:composite}) is having an RRL variable surrounded by a dusty disk,
which orbits and eclipses the star, causing the observed mean-magnitude changes. If the RRL star is single, then the disk is circumstellar, and the variations are caused by
inhomogeneities in the disk. In contrast, if the RRL is in a binary system, which is surrounded by a circumbinary disk, then part of the variability can be caused by
periodic obscuration by the disk as the RRL star is orbiting the center of mass of the system.

In this circumstellar disk case, the most probable source of dust is the RRL star itself.
Before evolving onto the HB, where they can be observed today, they previously passed through the red giant branch (RGB) phase,
losing $\sim0.2M_\odot$ material before the helium core flash \citep{2010A&A...517A..81G,2020MNRAS.498.5745T}.
At the same initial mass, age, and chemical composition, stars arriving onto the zero-age HB (ZAHB) bluer than the instability strip (IS)
must lose more mass on the RGB \citep{2009Ap&SS.320..261C}. Therefore, a larger fraction of longer-period RRL stars,
some of which are expected to be overluminous stars evolved off the blue HB, might have significant amounts of variable
detectable extinction from dust, if dust formed from material lost on the RGB is the cause of the mean-magnitude changes.
This possibility aligns well with the distribution of the sample of stars presented in the {\it top} and {\it middle} panels of Fig.~\ref{fig:statistics}.
On the other hand, in this RGB mass loss scenario, we could reasonably expect that the amount of dust around the star will slowly decrease
once it leaves the RGB tip, with the largest amount expected around pre-ZAHB stars. If these stars are pulsating as RRL variables
(i.e., falling within the IS), they are expected to show large negative period change rates, with the
most likely value being $\beta \sim -0.3\mathrm{\,d\,Myr}^{-1}$ \citep{2008A&A...489.1201S}.
In our sample, only two stars have relatively large (06316, 12833; $\beta \sim -0.3\mathrm{\,d\,Myr}^{-1}$) and one has a moderately large
(01081; $\beta \sim -0.2\mathrm{\,d\,Myr}^{-1}$) negative period change rate, with the vast majority having close to zero or slightly
positive rates. As approximately 1 star is expected to be in the pre-ZAHB phase for every 60 HB stars, the period changes
do not support this premise strongly \citep{2008A&A...489.1201S}.

Alternatively, dust in the stellar systems of RRL variables might be continually forming from ongoing
mass loss of the RRL stars themselves.
Mass loss has been observed in HB stars both redder \citep{2009AJ....138.1485D}, as well as bluer than RRL variables \citep{2016A&A...593A.101K}.
Continuous mass loss of RRL variables have been invoked with rates of up to $10^{-9}M_\odot\,\mathrm{yr}^{-1}$ in evolutionary calculations
to explain the mass dispersion of stars on the HB \citep{1994ApJ...423..380K}, while temporary mass loss in certain evolutionary phases
have been hypothesized to explain the period distribution of RRL variables in certain globular clusters \citep{2004ApJ...600..409C}.
Radial pulsation periodically decreases the surface gravity, which in turn might
lead to pulsation-induced mass loss \citep{1984Natur.312..429W}. Given the composition of the Sun \citep{2009ARA&A..47..481A}
scaled to a typical RRL iron abundance of $\mathrm{[Fe/H]}=-1.5$, the sum of the mass fractions of the main dust-forming elements (carbon, magnesium, silicon,
and iron) is $\sim0.016\%$. At the above quoted mass loss rate of $10^{-9}M_\odot\,\mathrm{yr}^{-1}$, an RRL star would lose
$\sim3.2\times 10^{17}\,\mathrm{kg}$ of these elements in a year. Supposing a dust formation efficiency of $1\%$, and a correction of a factor
by 100 due to the non-transiting parts of the dust, this mass loss would produce the estimated amount of dust
(see Section~\ref{sec:sed} and Appendix~\ref{sec:dustmass}) on a time scale of $\sim 10^{5}$\,yr.
Therefore, taking into account the total length of the HB phase (tens of Myr; \citealt{2008A&A...489.1201S}), this scenario might explain the
amount of observed dust, even if the mass loss is episodic, and relatively short-lived \citep{2004ApJ...600..409C}.
However, the mass loss rate of $10^{-9}M_\odot\,\mathrm{yr}^{-1}$ should probability be treated as an upper limit,
as larger rates would lead to significant changes in the morphology of the HB, leading to disagreement between HB models and observations \citep{1994ApJ...423..380K}.
Furthermore, we note that the fast winds inferred for red HB stars from chromospheric lines \citep{2009AJ....138.1485D} would
probably result in a low dust formation efficiency in the vicinity of the variables.

If the RRL is located in a binary system surrounded by a circumbinary disk, it is possible that the dust was
formed from material lost by the RRL variable, either in the preceding RGB phase or from mass loss on the HB, similarly to the circumstellar scenario discussed above.
While our knowledge of RRL binarity is very limited \citep{2021ApJ...915...50H}, presumably the most probable companions of RRL stars are
low-mass ($<1M_\odot$) main sequence stars and white dwarfs.
No significant mass loss can be expected from the former, while white dwarfs are known to have lost their
outer layers previously during the post-AGB phase of their evolution. Post-AGB stars in binary systems tend to form massive dusty
disks \citep{Beccari_Boffin_2019}, with masses of up to about $1\%$ of the Sun \citep{2021A&A...648A..93G}.
Therefore, what we observe today might be leftover material from this brief phase,
after which the post-AGB companion has evolved into the RRL variable that we observe today.
We can surmise that the amount of dust still in circumbinary orbit would mostly depend on the amount of dust formed in the post-AGB phase,
the fraction of the dust that survives photoevaporation once the post-AGB star becomes hotter before becoming a
white dwarf \citep{2003ARA&A..41..391V}, and the time elapsed since.
The latter can be quite short (tens of Myr) in systems where the two stars originally had very similar masses.
Solar-type stars in binaries around the Sun (RRL progenitors are old solar-type stars) have relatively large twin fractions
($\sim14\%$ for $2 < \log{P}{\rm [day]} < 6$; \citealt{2017ApJS..230...15M}), possibly supporting this scenario. If the post-AGB (now white dwarf) star
had a higher main sequence mass, the time elapsed since the post-AGB phase will be much longer (several Gyr), meaning
more time to reduce the amount of dust in the system.

Debris disks are common around white dwarf stars, and are thought to originate primarily from tidal disruption of planetary bodies \citep{2012ApJ...747..148D}.
Material from the disk is accreted onto the white dwarf, leading to detectable atmospheric pollution by metals, giving information
on the composition of planets \citep{2019AJ....158..242X}. A white dwarf in a binary system with an RRL star should be capable of breaking up minor bodies
(asteroids, comets, planetoids, and even planets) that enter its Roche radius.
While the dust from the minor body disruption is expected to settle into a debris disk around the white dwarf, if a fraction of it is expelled, it
could plausibly form a circumbinary disk.
The minor bodies might have formed during the second-generation planet formation process during
the post-AGB phase,
as suggested by similarities between transition disks and disks around post-AGB binaries \citep{2022A&A...658A..36K}.
The peculiar eclipse event shown by OGLE-BLG-RRLYR-09197 (Fig.~\ref{fig:09197}) is compatible with a white dwarf surrounded by a debris disk eclipsing the RRL variable in
the system, given the typical sizes of these objects. If confirmed, this would make this variable the only known eclipsing RRL star.

As a final note we remark that under this scenario, we expect most dust disks around RR~Lyrae stars not to be transiting their host stars due to their
specific inclinations, given their expected large orbital radii and relatively low scale heights.
Therefore, the true occurrence rate might be an order of magnitude larger than the $0.9\%$ incidence rate estimated in Section~\ref{sec:incidence}.

\subsection{Alternative 1: foreground interstellar matter variations} \label{sec:foreground}

Alternatively to being caused by circumstellar or circumbinary dust, a sufficiently clumpy foreground interstellar medium might
also be able to cause the observed mean-magnitude changes.
This effect was observed recently in the Galactic center region, caused by filaments of dust and gas under the influence of the
supermassive black hole in the center of the Milky Way \citep{2024AJ....168..166H}.
The large proper motion of stars in this region, combined with the relative thinness of these filaments, causes the variable extinction
effect. In the Milky Way, protostellar clumps and cores have typical sizes of $\sim1$\,pc and $\sim0.1$\,pc, respectively \citep{2007ARA&A..45..339B}.
If there is a proper motion difference between an RRL star and a protostellar clump or core in the foreground,
resulting in a perpendicular velocity difference of $100$\,km\,s$^{-1}$, in 10 years (the baseline of the OGLE-IV observations),
the RRL variable would pass behind $\sim0.001\,\mathrm{pc}$ ($\sim 200\,\mathrm{AU}$) of the foreground material.
If the clump or core has significant substructure below this scale, it might cause time-variable extinction, similar to the observed effect.
As protostellar clumps and cores are formed in dark clouds, they are strongly associated with their centers and the filaments connecting them
to other dark clouds (see Fig.~1 in \citealt{2007ARA&A..45..339B}, and Fig.~8 in \citealt{2015A&A...584A..92M}). These clouds and
filaments are well visible when they obscure the rich stellar background of the Galactic bulge. 
Therefore, we have inspected color-composite images from the Dark Energy Camera Plane Survey~2 \citep{2023ApJS..264...28S} and
the Pan-STARRS survey \citep{2016arXiv161205560C} for the regions around each of the 72 RRL stars with variable mean brightness.
The majority ($\sim2/3$) of these stars show no association with dark clouds, but the general patchy pattern of the foreground
interstellar material is virtually omnipresent. Less than 10 of them are located definitely behind dark clouds, typically near cloud edges
(the centers of most dark clouds generally cause enough extinction to make RRL stars undetectable by OGLE in the $V$ band).
The remaining stars are found near (less than 1\,arcminute away) the edges of dark clouds.

Additionally, we can estimate the number of dark clumps or cores that would be required to
obscure stars located in their background.
The final sample contains 72 stars located in 23 different OGLE-IV fields, each covering $\sim1.4$\,square degrees of the sky.
As the calculated occurrence rate is $\sim0.9\%$,  the obscuring material should cover $\sim0.29$\, square degrees of
these fields. We assume that typical clumps and cores are spherical and have diameters of $1$\,pc and $0.1$\,pc,
respectively \citep{2007ARA&A..45..339B}, and they are located at an average distance of $2$\,kpc
(i.e., they are in the foreground, relatively close to the Sun).
Then, the observed obscuration could be caused by $\sim 450$ clumps or, alternatively, $\sim45,\!000$ cores.
The APEX Telescope Large Area Survey of the Galaxy (ATLASGAL; \citealt{2018MNRAS.473.1059U}) identified and measured the properties of $\sim8000$
clumps in the Galactic disk, in an area $\sim15\times$ larger than the 23 OGLE fields (the region of $|b|<1.5^{\circ}$ and $5^{\circ}<|l|<60^{\circ}$).
Therefore, the required density of clumps is commensurable ($\sim80\%$) to those in the ATLASGAL fields. Taking into account several factors
(i.e., ATLASGAL fields are closer to the Galactic plane, while our stars are at larger Galactic latitudes;
some ATLASGAL sources are on the other side of the Galactic disk;
clumps and cores have different sizes;
on the other hand, some clumps and cores closer than 2\,kpc would appear much larger on the sky, etc.),
a few hundred clumps, together with a few thousand cores, might be able to cover the required sky area in these bulge fields.

We note that the above estimate implicitly assumes that substructures are present over the whole surface area of dark clumps and cores.
This is a very strong assumption, which to our knowledge lacks support in the literature.
Therefore, we conclude that this scenario, as compared to those discussed in Sect.~\ref{sec:insystem},
is less likely to be the reason behind the observed mean magnitude changes in bulge RR Lyrae stars.

\subsection{Alternative 2: intrinsic changes in RRL stars} \label{sec:intrinsic}

Our understanding of RRL variables is incomplete:
many of these stars show the Blazhko effect, a modulation of unknown origin \citep{2017MNRAS.466.2602P},
motivating the consideration of an unknown effect as the cause of the observed mean-magnitude variations.
The Blazhko effect modulates the light curves and pulsation periods of the variables, therefore it must be related to intrinsic processes.
In contrast, we find that the mean-magnitude changes are not affecting the light-curve shapes, amplitudes, or periods of the variables.
Furthermore, if mean-magnitude changes were caused by surface features, such as stellar spots, we would expect an additional modulation with the rotation
period of the stars, as is routinely observed in spotted stars \citep{2019ApJ...879..114I}. RRL variables are slow rotators with rotational periods
of at least 40\,days\footnote{Calculated from the spectroscopically measured upper limit on $V_\mathrm{rot}\sin(i)<5$\,km\,s$^{-1}$ \citep{2019AJ....157..153P} for typical RRL radii.}.
We detect no additional periodicities in our sample between 40 and 180\,days, which is at odds with the expected presence of stellar spots in this scenario.
Some variables show annual trends in the OGLE-IV photometry, but these are instrumental in nature.
Furthermore, this scenario is also strongly challenged by the eclipse event shown by 09197. It is unlikely that stellar spots could grow to diminish
the light of the star by $\sim20\%$, then disappear during the $\sim12$\,day time span of the fading episode.
Consequently, we do not think this is a plausible scenario for explaining the observed mean-magnitude change phenomenon.

\subsection{Future outlook}

Further observations of these RRL variables are necessary to determine the true cause of the observed mean-magnitude changes.
Continued photometric monitoring is required to determine if the signatures are periodic,
which would put strong constraints on its possible causes.
If the changes are caused by dust obscuration in the stellar system hosting the RRL variable (Sect.~\ref{sec:insystem}),
or intrinsic processes (Sect.~\ref{sec:intrinsic}), then closer, brighter RRL stars with variable mean magnitudes should also exist.
Conversely, if they are caused by variable interstellar extinction (Section~\ref{sec:foreground}), it should be widespread among bulge stars,
and readily identifiable in existing microlensing survey data.

Besides finding more RRL stars showing the same kind of mean-magnitude variability, the variables reported here
should be subject to further photometric, spectroscopic, and even polarimetric studies.
Multi-band observations can be used to determine the shape of the extinction curve produced by the dust, thereby constraining its
properties \citep{2023ApJ...957L..28M}. As RRL stars are old, metal-poor objects, the properties of dust around them might give unique
information on dust chemistry at low metallicities.
Furthermore, for variables with the highest extinction values, the flux in the bluer bands can be almost completely
absorbed, and what is observed is the scattered light from dust transiting outside the stellar disk. This can lead to blueing of the colors
in ultraviolet-optical bands bluer than the $V$ band, as observed for the young, solar-like star ASASSN-21qj \citep{2023Natur.622..251K}.
The polarization properties of light are modified by intervening dust, both in the interstellar \citep{2015ARA&A..53..501A} and
circumstellar \citep{2017ApJ...839...56T} media.
Consequently, differential measurements between bright and faint periods of the same stars can in principle separate the two signals,
further constraining the dust parameters.

Spectroscopy will be crucial for studying these objects. Radial velocities, combined with proper motions and distances, will provide the kinematical
information \citep{2020AJ....159..270K} needed to correctly assign each star to its host stellar population (Galactic bulge, disk, or halo). 
Post-AGB stars with disks generally show a depletion of refractory elements in their spectra, caused by the accretion of material from the
circumbinary disk \citep{2019A&A...629A..49O}. If the post-AGB companion is now the RRL we are observing, its chemical abundance pattern might show a similar
distribution. The peculiar abundance pattern of the RRL star TY~Gru \citep{2006AJ....132.1714P}, showing carbon and neutron-capture element overabundances, can
be explained by accretion of material from an AGB companion \citep{2013MNRAS.435..698S}. The same scenario can also account for the existence of
carbon-enhanced metal-poor RRL variables \citep{2014ApJ...787....6K}. Transmission spectroscopy might also reveal the presence of gas in the disk, similarly
to its frequent detection in debris disks transiting white dwarfs \citep{2020ApJ...905...56M}.

In the obscuration by molecular clumps and cores scenario (Sect.~\ref{sec:foreground}), however, different observables are expected. With no circumstellar material,
the SED would reveal no mid-IR excess. However, strong mm and sub-mm emission from the cold foreground material, as well as emission from molecules,
particularly CO \citep{2007ARA&A..45..339B}, would be expected to coincide with the positions of the RRL showing
mean-magnitude changes.

As a final consideration, we note that we cannot rule out that more than one proposed scenario contributes to the observed mean-magnitude changes in
the RRL variables in the Galactic bulge fields. Furthermore, the source of the dust might be different from any of our proposed scenarios.
Therefore, we encourage observational follow-up to reveal the true nature of the mean-magnitude changes in RRL variables.

\section*{Data availability}
Full Figure~\ref{fig1} is available on Zenodo via \url{https://zenodo.org/records/17953774}.
Full Table~\ref{tab:properties} is available at the CDS via \url{asdasdasd}.
Table~\ref{tab:properties} is only available in electronic form at the CDS via anonymous ftp to
cdsarc.u-strasbg.fr (130.79.128.5) or via http://cdsweb.u-strasbg.fr/cgi-bin/qcat?J/A+A/.

\begin{acknowledgements}
      The research leading to these results has received funding from the European Research Council (ERC) under the European Union's
      Horizon 2020 research and innovation programme (grant agreement No. 695099).
      This work has been funded by the National Science Centre, Poland, grant no.~2022/45/B/ST9/00243
      and grant 2024/WK/02 of the Polish Ministry of Science and Higher Education.
      Support for M.C. is provided by ANID's FONDECYT Regular grant \#1231637; ANID's Millennium Science Initiative through grants
      ICN12\textunderscore 009 and AIM23-0001, awarded to the Millennium Institute of Astrophysics (MAS); and ANID's Basal project FB210003.
      This research has made use of the KMTNet system operated by the Korea Astronomy and Space Science Institute (KASI) at three host sites
      of CTIO in Chile, SAAO in South Africa, and SSO in Australia. Data transfer from the host site to KASI was supported by the
      Korea Research Environment Open NETwork (KREONET).
      This paper utilizes public domain data obtained by the MACHO Project, jointly funded by the
      US Department of Energy through the University of California, Lawrence Livermore National Laboratory under
      contract No. W-7405-Eng-48, by the National Science Foundation through the Center for Particle Astrophysics
      of the University of California under cooperative agreement AST-8809616, and by the Mount Stromlo and
      Siding Spring Observatory, part of the Australian National University.
      This paper makes use of data obtained by the MOA collaboration with the 1.8-metre MOA-II telescope at the
      University of Canterbury Mount John Observatory, Lake Tekapo, New Zealand.
      The MOA collaboration is supported by JSPS KAKENHI grant and the Royal Society of New Zealand Marsden Fund.
      These data are made available using services at the NASA Exoplanet Archive, which is operated by the
      California Institute of Technology, under contract with the National Aeronautics and Space Administration
      under the Exoplanet Exploration Program.
      The authors are grateful to Jean-Baptiste Marquette for providing the EROS-2 data.
      This research has made use of the SVO Filter Profile Service ``Carlos Rodrigo", funded by MCIN/AEI/10.13039/501100011033/ through grant PID2020-112949GB-I00.
\end{acknowledgements}

\bibliographystyle{aa}
\bibliography{arxiv}

\begin{appendix}

\section{Properties of the optical photometry}\label{sec:data2}

The optical light curves used during our analysis originate from a number of different surveys, most of which have photometric
systems different from that of OGLE-IV. Here, we discuss these differences and the reduction/transformation of some of the
data onto a magnitude scale.

The OGLE-III data were obtained with the same telescope as the OGLE-IV data, but using the second generation OGLE camera \citep{2003AcA....53..291U},
and different filters, but both are calibrated to the $V$ and $I$ filters of the Johnson-Kron-Cousins system.
The MACHO survey RRL light curves \citep{1998ApJ...492..190A}, obtained using two custom wideband filters,
were downloaded from the MACHO TAP service\footnote{\url{https://macho.nci.org.au/}}, using {\tt TOPCAT} \citep{2005ASPC..347...29T}.
The EROS-2 observations were obtained in two custom bands simultaneously. The photometry
was obtained using the AstroMatic software suite \citep{1996A&AS..117..393B,2006ASPC..351..112B,2011ASPC..442..435B}, and is on a relative,
uncalibrated magnitude scale. The reduction was performed and the light curves were kindly provided by Jean-Baptiste Marquette.
The KMTNet observations were obtained in the $I$ band with each of the identical 1.6m KMTNet telescopes at three different observatories, 
KMTNet-CTIO in Chile, KMTNet-SAAO in South Africa, and KMTNet-SSO in Australia \citep{2016JKAS...49...37K}.
The raw, 800MB images from each of the 4-square degree field mosaic imagers are first saved on local disks at each observatory. 
These are transferred to the Korean data center through the Internet as soon as data acquisition is completed.
Raw images of the Galactic bulge fields are pre-processed and divided into 256 stamps for difference image analysis.
A pre-defined catalog of sources is used to extract light curves on a differential flux scale.
Generally, two constants are needed to tie differential fluxes to the magnitude scale. These are the flux level of the
source on the reference image, and its corresponding magnitude value.
The KMTNet observations are mostly contemporary with the OGLE-IV data (starting in 2016), therefore,
the differential flux measurements were transformed to the magnitude scale by using the light-curve solution of the OGLE-IV
$I$-band data. We note that as this transformation involves the determination of only these two parameters, 
it is not capable of artificially introducing mean-magnitude changes, if they are not already present in the data.
The MOA observations are similarly available as differential flux measurements at the
NASA Exoplanet Archive\footnote{\url{https://exoplanetarchive.ipac.caltech.edu/docs/MOAMission.html}};
however, they were obtained using a custom wide-band red filter.
As the effective wavelength of this filter is roughly halfway between the Johnson-Kron-Cousins $R$ and $I$ filters, we decided to
transform the differential fluxes onto an arbitrary magnitude scale by requiring the RRL stars to have $\sim10\%$ larger
pulsation amplitudes in the MOA survey photometry than in the OGLE $I$-band data (based on a linear interpolation of the effective wavelengths
of the $V$ and $I$ bands to that of the MOA filter, and using a $V$-to-$I$ amplitude ratio of $\sim1.5$ for RRab stars).

\section{Fourier fitting of light curves with changing mean magnitudes}\label{sec:fourier}

RRL light curves are traditionally fit using a truncated Fourier series, which in its linear form for order $\mathcal{F}$ can be written as:

\begin{equation}
    m = m_0 + \sum_{k=1}^{\mathcal{F}} \left[ A_k \sin(k \omega t) + B_k \cos(k \omega t) \right] , \label{eq:fourier}
\end{equation}

\noindent where $m_0$ is the mean magnitude, $\omega = 2\pi / P$ is the angular frequency for period $P$, $t$ is
time, $A_k$ and $B_k$ are the $k$-th order Fourier coefficients.
As this is a linear equation, it can be solved with the ordinary least squares (OLS) method for the $A_k$ and $B_k$ coefficients.
For a time series containing $i$ points at times $t_i$, the design matrix will have the form:

\begin{equation}
    X = 
    \begin{bmatrix}
        1 & \sin(\omega t_1) & \cos(\omega t_1) & \ldots & \sin(\mathcal{F}\omega t_1) & \cos(\mathcal{F}\omega t_1) \\
        1 & \sin(\omega t_2) & \cos(\omega t_2) & \ldots & \sin(\mathcal{F}\omega t_2) & \cos(\mathcal{F}\omega t_2) \\
        \vdots & \vdots      & \vdots           & \ddots & \vdots &\vdots\\
        1 & \sin(\omega t_i) & \cos(\omega t_i) & \ldots & \sin(\mathcal{F}\omega t_i) & \cos(\mathcal{F}\omega t_i)
    \end{bmatrix}, \label{eq:x1}
\end{equation}

\noindent where the first column, filled with ones, is called the bias term and corresponds to the mean magnitude $m_0$ in Equation~\ref{eq:fourier}.
In order to allow the mean magnitude to change while keeping the light curve shapes intact, we replace the bias term with
coefficients for a continuous linear piecewise regression. In this case, the coefficients are linearly scaled between the closest
pre-defined breakpoints $t_{z}$ (here $z$ is an integer whose value is between 1 and $N_\mathrm{bp}$, the total number of breakpoints,
and $t_z < t_{z+1}$ strictly for all $z$ values):

\begin{equation}
    f_{z,i}(t_i) = \begin{cases}
        \frac{t_i-t_z}{t_{z+1}-t_z}, & \mbox{if } t_z \leq t_i \leq t_{z+1}\\
        \frac{t_z-t_i}{t_z-t_{z-1}}, & \mbox{if } t_{z-1} \leq t_i \leq t_z\\
        0 & \mbox{otherwise},
    \end{cases} \label{eq:sawtooth}
\end{equation}

\begin{table*}[ht!]
\caption{\label{tab:surveys}Information on the RR~Lyrae stars from other bulge microlensing surveys}
\centering
\begin{tabular}{@{\extracolsep{\fill}}c@{\extracolsep{1.0em}}c@{\extracolsep{0.6em}}c@{\extracolsep{0.6em}}c@{\extracolsep{0.6em}}c@{\extracolsep{0.6em}}c|
@{\extracolsep{1.5em}}c@{\extracolsep{1.0em}}c@{\extracolsep{0.6em}}c@{\extracolsep{0.6em}}c@{\extracolsep{0.6em}}c@{\extracolsep{0.6em}}c|
@{\extracolsep{1.5em}}c@{\extracolsep{1.0em}}c@{\extracolsep{0.6em}}c@{\extracolsep{0.6em}}c@{\extracolsep{0.6em}}c@{\extracolsep{0.6em}}c@{\extracolsep{\fill}}}
\hline\hline
ID     &  MA  & E2  & O3  & MO  & KM  & ID      & MA  & E2  & O3  & MO  & KM  & ID      & MA  & E2  & O3  & MO  & KM   \\
\hline
00663  &  --  &  I  &  I  & --  &  I  &  08281  &  I  &  I  &  I  &  S  &  S  &  12586  &  S  & --  &  S  &  S  &  I   \\
00835  &  --  &  I  &  I  & --  &  I  &  08650  & --  &  S  &  I  &  S  &  I  &  12605  & --  &  I  &  S  &  S  &  S   \\
00853  &  --  &  I  &  I  & --  &  I  &  08752  &  I  & --  &  S  &  S  &  S  &  12793  & --  &  S  &  S  &  I  & --   \\
00942  &  --  &  I  &  S  & --  &  I  &  09024  & --  &  I  &  I  &  S  &  I  &  12833  & --  & --  &  S  & --  &  S   \\
01081  &  --  &  I  &  S  & --  &  S  &  09197  &  S  & --  &  S  &  S  &  I  &  12843  &  I  & --  &  S  &  I  &  I   \\
01096  &  --  &  S  &  I  & --  & --  &  09247  & --  & --  &  S  &  S  &  S  &  13260  &  I  &  S  &  S  & --  &  I   \\
01591  &  --  &  S  &  S  & --  &  S  &  09893  & --  & --  &  S  & --  &  I  &  13433  &  I  & --  &  S  &  S  &  I   \\
02401  &  --  &  S  &  S  &  S  &  S  &  10082  & --  &  I  &  I  &  S  &  I  &  13720  & --  &  S  &  S  &  S  &  S   \\
04219  &  --  &  S  &  S  & --  & --  &  10084  &  I  &  I  &  I  &  S  &  S  &  13914  &  I  & --  &  S  &  S  &  S   \\
04353  &  --  & --  &  S  & --  &  I  &  10745  &  I  &  I  &  S  & --  &  I  &  14475  &  I  & --  &  S  &  S  &  I   \\
04444  &  --  & --  &  I  & --  &  S  &  10751  & --  & --  &  S  &  S  &  I  &  14947  &  I  &  I  &  S  & --  &  I   \\
04501  &  --  & --  &  S  &  S  &  S  &  10782  & --  & --  &  S  &  S  &  S  &  15075  & --  &  I  &  I  &  S  &  S   \\
04871  &  --  & --  &  S  &  S  &  S  &  10812  &  S  &  I  &  I  &  I  &  I  &  15304  & --  & --  &  S  &  S  &  S   \\
05016  &  --  & --  &  S  & --  & --  &  10834  & --  &  S  &  S  & --  & --  &  16194  &  I  & --  &  I  & --  &  S   \\
05217  &  --  &  I  &  S  & --  &  S  &  10994  &  I  &  I  &  S  &  S  &  I  &  16229  &  I  &  I  &  I  &  S  &  S   \\
05572  &  --  &  I  &  S  &  S  &  S  &  11061  & --  & --  &  S  &  S  &  S  &  21903  & --  &  I  & --  & --  &  S   \\
06019  &  --  & --  &  S  &  S  &  S  &  11608  &  S  & --  &  S  &  S  &  I  &  22755  & --  & --  & --  & --  & --   \\
06165  &  --  &  I  &  S  &  S  &  S  &  11748  &  I  &  I  &  I  &  S  &  I  &  23406  & --  & --  & --  & --  &  S   \\
06316  &  --  &  I  &  S  &  S  &  S  &  11813  & --  & --  &  S  & --  &  S  &  29180  & --  &  S  & --  & --  &  I   \\
07208  &   I  & --  &  S  &  S  &  I  &  11931  & --  & --  &  S  &  S  &  I  &  31258  & --  &  S  & --  &  S  &  S   \\
07219  &  --  & --  &  I  & --  &  S  &  12056  & --  & --  &  S  &  S  &  S  &  31664  & --  & --  & --  & --  &  I   \\
07830  &  --  & --  &  I  &  S  &  I  &  12157  & --  &  S  &  S  &  S  &  S  &  32226  &  S  &  S  & --  & --  &  I   \\
07891  &  --  & --  &  S  &  S  &  S  &  12237  &  S  &  S  &  S  &  S  & --  &  33665  & --  & --  & --  &  S  &  S   \\
08215  &  --  &  I  &  S  &  S  &  I  &  12429  & --  & --  &  S  &  S  &  S  &  34373  &  S  & --  &  S  &  S  &  S   \\
\hline
\end{tabular}
\tablefoot{The first, seventh and thirteenth columns provide the OGLE identifiers in the format OGLE-BLG-RRLYR-$ID$.
The following five columns give information about the presence of the long-term mean-magnitude changes in the  
MACHO, EROS-2, OGLE-III, MOA, and KMTNet surveys, respectively. Stars marked with ``--''
have no photometry available in a given survey. Stars marked with ``S'' have photometry supporting the presence of
long-term magnitude changes, while ``I'' denotes variables where the photometry is inconclusive, in most cases because
of the quality and/or quantity of the data in a given survey for that particular star.
}
\end{table*}

\noindent where $f_{z,i}(t_i)$ are the piecewise regression coefficients.
It can be seen trivially that $\sum_{n=1}^{N_\mathrm{bp}} f_{n,i}(t_i) = 1$ for all $i$ values when $t_{\mathrm{bp},0} \leq \min(t_i)$
and $ \max(t_i) \leq t_{\mathrm{bp},z}$. In practice, we assign $N_{\mathrm{bp},s}$ breakpoints to $s$ seasons of observations, and align the first and last
breakpoints to the beginning and end of each season. Therefore, the total number of breakpoints we have for a given light curve is
$N_\mathrm{bp} = s N_{\mathrm{bp},s}$. When the $f_{z,i}$ values are substituted in place of the bias term in the design matrix shown in Equation~\ref{eq:x1},
and OLS is used to solve the problem, the corresponding coefficients give the momentary mean-magnitude values $m_{0,z}$ for each time $t_z$.
In our implementation, we use the {\tt LinearRegression} class of the {\tt linear\_model} module of the {\tt scikit-learn} Python
package \citep{scikit-learn} to solve the OLS equation.

RRL variables in the Galactic bulge area of the OGLE-IV survey in most cases have several thousands of points in the $I$ band, allowing accurate
fitting of the light-curve shape with a high-order Fourier series (Equation~\ref{eq:fourier}), and many breakpoints to model the change in the mean magnitudes.
In contrast, in our final sample the $V$-band data have at most 230 points, meaning they cannot be analyzed the same way as the $I$-band light
curves. Therefore, a different approach is used: the shape of the mean-magnitude changes of the $I$-band light curves are adopted for the $V$ band
as well, but allowing their amplitudes to change.
This is done by calculating the mean-magnitude change relative to the median of the breakpoints in the $I$-band for each $V$-band data point.
We add these values to the design matrix shown in Equation~\ref{eq:x1} as an extra column. In effect, this increases the number of parameters by one,
the additional coefficient representing the amplitude ratio $A_V/A_I$ between the $V$ and $I$ bands,
as inferred from the mean-magnitude changes in these two bandpasses.

A typical issue in fitting light curves with truncated Fourier series is the determination of the optimal Fourier order $\mathcal{F}$.
In our case, this is compounded with the addition of $N_\mathrm{bp}$ breakpoints as extra parameters. The number of breakpoints
per season was set for every star individually, depending on the complexity of the mean-magnitude changes affecting the
light curve.
The optimal Fourier order was determined by calculating the AIC value corrected for finite samples \citep{CAVANAUGH1997201} for
Fourier orders between 5 and 30 for the $I$ band, and
1 and 30 for the $V$ band. We noticed that in some cases, adopting the solution with the lowest AIC value would result in slightly
overfit light-curve shapes, while lower orders, with only slightly higher AIC values, would produce more reasonable fits.
Therefore, in every case we choose the lowest order with an AIC value not larger than 10 compared to
the one with the lowest AIC value, instead of simply adopting the latter.

Changes in the pulsation phases of the variables with respect to the adopted pulsation period present a further complication
for obtaining an accurate fit of the light curves. These can be either due to binarity or because of intrinsic changes in the
pulsation period of the RRL star itself.
For stars with a binary Observed minus Calculated ($O-C$) solution available in our study on RRL binarity \citep{2021ApJ...915...50H}, we subtract it
from the timings of the light curve data. 
For the remaining stars, we construct their $O-C$ diagrams as in our previous study, but using an initial fit of their mean-magnitude
changes to remove the latter from the $V$- and $I$-band light curves.
For variables with no apparent period change over the time base of observations, these are used only to iterate and obtain a
more accurate pulsation period.
For stars with linear period changes, these are fit using a parabola, and subtracted from the timings of the light curves.
After correcting the timings of the observations, we determine the final light-curve parameters by applying again our modified truncated
Fourier series method, as described above.
One additional star was found to be a strong binary candidate during the analysis, whose special treatment is
described in Appendix~\ref{sec:newbin}.

\section{Additional information on the light curves of the variables on the final list}\label{sec:otherinfo}

We provide information on the presence of the mean-magnitude changes in the bulge microlensing survey light curves other than OGLE-IV for the stars on the final list
in Table~\ref{tab:surveys}.
Whether these are detected for a star in a particular survey depends on their amplitude and the
quality of the photometry in the survey.
As shown by the comparison between light curves from different surveys in Figure~\ref{fig:composite},
the early MACHO and EROS-2 observations have the lowest quality in general, but the change in the mean magnitudes can still be ascertained by them for
some of the stars. The OGLE-III light curves show the same kind of behavior as the OGLE-IV data, even for stars which were only observed for
a few seasons. While individual MOA data points are generally less accurate than OGLE data,
due to their large quantity, the MOA data still clearly reveal the same light curve behavior as seen in the OGLE-III and IV data for the overlapping seasons.
The KMTNet and OGLE-IV data are also in general agreement.

The analysis also revealed that 10 RRab variables in the final sample show the Blazhko effect in their light curves. While Fourier series-based modeling of the light curve would ideally include
extra terms to account for this, the limited number of $V$-band light curve points prevents a consistent analysis. Therefore, we only
fit the average Fourier terms, and note the estimated modulation periods in Table~\ref{tab:Blazhko}.

\begin{table}[]
\caption{\label{tab:Blazhko}Variables with Blazhko effect in the sample}
\centering
\begin{tabular}{cr@{}lc}
\hline\hline
ID                  &    \multicolumn{2}{r}{$P_\mathrm{Bl,1}$}  & $P_\mathrm{Bl,2}$ \\
                    &    \multicolumn{2}{r}{(days)}  & (days) \\\hline
\,00663\,               &    44        & .2  & --      \\
\,00853\,               &    24        & .34 & 29.16   \\
\,01096\,               &    61        & .76 & --      \\
\,05572\,               &    66        & .21 & --      \\
\,06316\,               &    20        & .98 & --      \\
\,08650\,               &    24        & .07 & --      \\
\,08752\,               &    27        & .30 & 51.57   \\
\,09197\,               &    53        & .45 & --      \\
\,11748\,               &    37        & .64 & --      \\
\,12843\,               &    $\sim380$ &     & --      \\
\hline
\end{tabular}
\tablefoot{The first column provides the OGLE identifier in the format OGLE-BLG-RRLYR-$ID$.
The second and third columns give the periods of the primary and secondary modulations (if present), respectively.
}
\end{table}

\section{Mass estimate for the transiting circumstellar dust}\label{sec:dustmass}

We estimate a lower limit for the mass of the dust transiting the three stars in our sample with the highest
extinction amplitudes, while still having reliable $r_{I,V}$ estimates, i.e., for OGLE-BLG-RRLYR-05572, -11061, and -34373
(shown on the {\it top-left}, {\it bottom-left} and {\it bottom-right} panels of Figure~\ref{fig:composite}, respectively).
Assuming that the obscuring material is on a circular orbit around the RRL variable at a distance of 10\,AU, a typical RRL
mass of $0.65\,M_\odot$ results in a transversal velocity of $\sim7.6$\,km\,s$^{-1}$.
Adopting a recent period-radius relation measured from Baade-Wesselink analysis of RRab stars \citep{2024A&A...684A.126B},
we estimate the stellar radii as $\sim5.65\,R_\odot$, $\sim5.46\,R_\odot$ and $\sim4.38\,R_\odot$ for the three RRL stars,
respectively (for the last variable, an RRc star, after its first overtone period is converted to the fundamental one using
$\log P_\mathrm{F} = \log P_\mathrm{1O}+0.127$; \citealt{1971PASP...83..697I}).
Dividing the stellar diameters by the transversal velocity gives the crossing time for the dust, which is
$\sim12.0$\,d, $\sim11.6$\,d, and $\sim9.3$\,d, respectively. We note that these values are very comparable to the
$\sim12$\,d length of the short dimming event shown in Fig.~\ref{fig:09197} for OGLE-BLG-RRLYR-09197.
We calculate the total amount of extinction in the $I$ band during the time base of the OGLE-IV observations
as the integral between the 10th percentile of the magnitude values of the breakpoints ($m_{0,z}$) and the interpolated
mean-magnitude values between each of the breakpoints (in units of $\mathrm{mag\times day}$).
The total amount of the $V$-band extinction is calculated by dividing this value by the $r_{I,V}$ value measured for a given star.
Then, further dividing this by the corresponding crossing times,
we get an estimate of the total amount of extinction caused by the transiting material.
For the three considered variables, these are
$\sim74$\,mag, $\sim42$\,mag, and $\sim111$\,mag, respectively.
To convert these into masses, we need to adopt a dust mass-to-extinction ratio, for which we suppose that the transiting dust has similar properties
to that of the interstellar dust.
As the dust mass-to-extinction ratio is not estimated directly, but with respect to hydrogen, we calculate it through
the hydrogen-to-extinction and gas-to-dust ratios. Specifically, we adopt these in the form \citep{2017MNRAS.471.3494Z}:
$N_\mathrm{H}(\mathrm{H\,cm}^2) = (2.08\pm0.02)\times 10^{21} A_V \mathrm{(mag)}$ and
$M_\mathrm{gas}/M_\mathrm{dust}\approx140$, where $N_\mathrm{H}$ is the hydrogen column density.
The gas mass is higher than the hydrogen mass by a factor of 1.4 due to the
contribution of helium. Taking this into account, and combining these two relations, the dust mass-to-extinction ratio is:
\begin{equation}
    \label{eq:dust}
    M_\mathrm{dust} (\mathrm{g}\,\mathrm{cm}^{-2}) \approx 3.5 \times 10^{-5} A_V \mathrm{(mag)}.
\end{equation}
Multiplying the area of the stellar disk
by the total amount of extinction
and combining with Eq.~\ref{eq:dust}
gives mass estimates of $\sim1.3\times10^{18}$\,kg, $\sim6.7\times10^{17}$\,kg, and $\sim1.1\times10^{18}$\,kg,
for OGLE-BLG-RRLYR-05572, -11061, and -34373, respectively.

It must be emphasized that the calculation presented here is only a lower estimate for the mass of the transiting material for these variables.
Most reasonable modifications of the employed assumptions (e.g., smaller orbit, material transiting above and below the stellar disk, inhomogeneities, etc.)
further increase the estimated masses. It is clear from Fig.~\ref{fig:composite} that the material has not yet completed a full orbit around
either one of the stars considered here, meaning that in the circumstellar dust scenario, its orbit must have $a\gtrsim4$\,AU.
Furthermore, it is not clear from the current data whether any of our stars has been observed without the
additional extinction from the circumstellar material at all during the baseline of OGLE-IV observations, meaning that
the mass for the transiting dust might be significantly more than the calculation here suggests.

\begin{figure}[t!]
    \centering
    \includegraphics[width=\hsize]{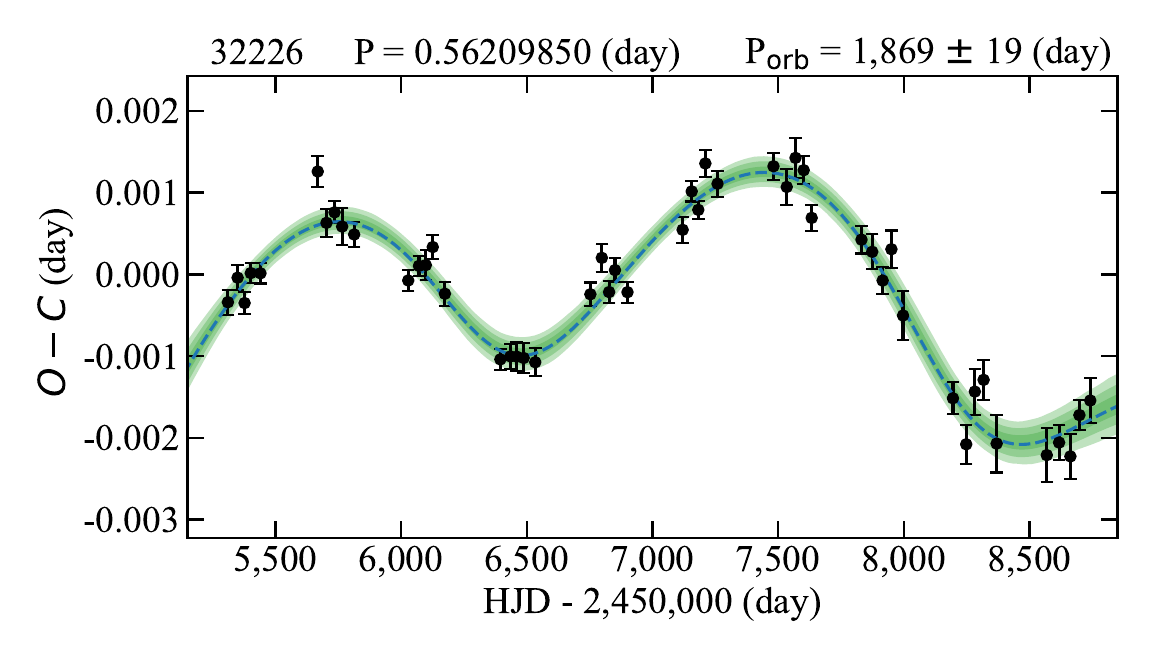}
    \caption{$O-C$ diagram of OGLE-BLG-RRLYR-32226.
    Black points show the measured $O-C$ values. The dashed line shows the average solution from the
    MCMC modeling \citep{2021ApJ...915...50H}. The green shaded regions denote the 1, 2, and 3$\sigma$
    credible intervals of the solutions. Above the panel, the OGLE ID, the pulsation and orbital periods
    are given.}
    \label{fig:32226}
\end{figure}
                                                       
\section{The binary parameters of OGLE-BLG-RRLYR-32226}\label{sec:newbin}

During our analysis, we noticed that the $O-C$ diagram of OGLE-BLG-RRLYR-32226 shows the characteristic periodic
oscillation pattern expected from variables located in a binary system due to the LTTE.
In fact, the binarity of this variable was already suspected during our dedicated RRL binary search \citep{2021ApJ...915...50H},
but the mean-magnitude changes were mistaken for the Blazhko effect, and it was discarded from the list of binary candidates.
A more careful analysis has revealed that the light curve of the star is stable, but it is apparently affected by the
LTTE and the mean-magnitude changes simultaneously, both at a relatively small amplitude. Therefore, an iterative procedure was
adopted for the analysis of this star, in order to decouple these two effects in the light curve.

First, we performed an $O-C$ analysis on the original light curve, following the Bayesian, Markov Chain Monte Carlo (MCMC) based procedure
outlined in our binarity study \citep{2021ApJ...915...50H}.
The timings of the light curve points were corrected with this initial $O-C$ solution, and
the resulting data were analyzed for the change in mean magnitudes as described above.
Then, we repeated the $O-C$ analysis after correcting the original light curve for the change in mean magnitudes.
This solution is shown in Figure~\ref{fig:32226}, and the derived parameters are listed in Table~\ref{tab:32226}.
These are: the orbital period ($P_\mathrm{orb}$), the projected semimajor axis ($a_1 \, \sin(i)$), the eccentricity ($e$), the argument of pericenter ($\omega$),
the pericenter passage time ($T_0$), as well as the pulsational period change rate $\beta$.
From these, we also derive the following quantities: the orbital radial velocity semi-amplitude ($K_1$), the mass function ($f(m)$) and
the minimum companion mass ($M_\mathrm{S,min}$), and also list them in Table~\ref{tab:32226}.
Finally, we corrected the timings of the light curve with this second $O-C$ solution, and re-analyzed the mean-magnitude changes,
with the results of this final fit being listed in Table~\ref{tab:32226}.

\begin{table}[t!]
\caption{\label{tab:32226}Binary parameters of OGLE-BLG-RRLYR-32226}
\centering
\begin{tabular}{@{}l@{}r@{}c@{}l@{}}
\hline\hline
Fit parameters                        &    Value & & $68\%$ credible interval \\
\hline
$P_\mathrm{orb}$                      &     1869 &\,$\pm$\,& 19\,days                 \\
$a_1 \sin i$                          &    0.202 &\,$\pm$\,& 0.007\,AU                \\
$e$                                   &    0.222 &\,$\pm$\,& 0.064                    \\
$\omega$                              &     -126 &\,$\pm$\,& 16\,deg                  \\
$T_0$\footnotemark[1]                 &     6348 &\,$\pm$\,& 87\,days \\
$\beta$                               &   -0.214 &\,$\pm$\,& 0.010\,day\,Myr$^{-1}$   \\
\hline
Derived parameters                    &                                           \\
\hline
$K_1$                                 &     1.21 &\,$\pm$\,& 0.05\,km\,s$^{-1}$       \\
$f(m)$                                &  0.00031 &\,$\pm$\,& 0.00003\,$M_\odot$      \\
$M_{\mathrm{S,min}}$\footnotemark[2] &   0.0538 &\,$M_\odot$          \\
\hline
\end{tabular}
\tablefoot{{Parameters derived from the MCMC fit of the $O-C$ diagram shown in Fig.~\ref{fig:32226}}.\\
\tablefoottext{1}{HJD$-$2,450,000.}
\tablefoottext{2}{Derived assuming $m_1\equiv0.65M_\odot$ and $i\equiv90^\circ$.}
}
\end{table}

\end{appendix}
\end{document}